\documentclass[11pt,superscriptaddress]{article}

\usepackage{geometry}
\usepackage{amssymb,color,paralist,amsmath,epsfig}
\usepackage{graphicx}
\usepackage{comment}
\usepackage{amsfonts}
\usepackage{relsize}
\usepackage{nicefrac,esint}
\usepackage{float}
\usepackage{mathrsfs}
\usepackage{amsthm}
\usepackage{caption} 
\usepackage[affil-it, auth-sc]{authblk}
\usepackage[utf8]{inputenc}
\usepackage{cite}

\usepackage{eso-pic}

\usepackage{color, soul}
\usepackage{verbatim}
\usepackage{lineno}
\usepackage{todonotes}

\makeatletter
\newcommand\footnoteref[1]{\protected@xdef\@thefnmark{\ref{#1}}\@footnotemark}
\makeatother

\presetkeys{todonotes}{inline, color=blue!30, size=\small}{}

\definecolor{darkred}{rgb}{0.9, 0.0, 0.0}
\definecolor{darkgreen}{rgb}{0.0, 0.5, 0.0}
\usepackage[colorlinks,bookmarks,linkcolor=darkred, citecolor=darkgreen]{hyperref}

\newcommand{\beq}{\begin{equation}}
\newcommand{\eeq}{\end{equation}}

\newcommand{\ber}{\begin{eqnarray}}
\newcommand{\eer}{\end{eqnarray}}

\newcommand{\nl}{\nonumber \\ }
\newcommand{\order}{{\cal O}}

\newcommand{\myxi}{r_H}
\newcommand{\mytau}{r_t}
\newcommand{\GF}{\mathrm{G}_{\mathrm{F}}}

\newcommand{\mynf}{n_f}

\newcommand{\rL}{\mathrm{L}}
\newcommand{\rR}{\mathrm{R}}

\allowdisplaybreaks

\begin{document}


\AddToShipoutPictureFG*{
    \AtPageUpperLeft{\put(-60,-60){\makebox[\paperwidth][r]{FERMILAB-PUB-19-559-T}}}  
    }%

\title{On the effective theory of neutrino-electron and neutrino-quark interactions}

\author{Richard J. Hill}
\author{Oleksandr Tomalak}

\affil{Department of Physics and Astronomy, University of Kentucky, Lexington, KY 40506, USA}
\affil{Theoretical Physics Department, Fermi National Accelerator Laboratory, Batavia, IL 60510, USA }

\date\today

\maketitle

\begin{abstract}
  We determine the four-fermion effective theory of neutrino interactions within the
  Standard Model including one-loop electroweak radiative corrections, in
  combination with the measured muon lifetime and precision electroweak data.
  Including two-loop matching and three-loop running corrections, we determine lepton
  coefficients accounting for all large logarithms through relative order $\order(\alpha \alpha_s)$
  and quark coefficients accounting for all large logarithms through $\order(\alpha)$.
  We present four-fermion coefficients valid in $\mynf=3$ and $\mynf=4$ flavor quark theories,  
  as well as in the extreme low-energy limit. We relate the coefficients in this limit to neutrino charge radii 
  governing matter effects via forward neutrino scattering on charged particles.
\end{abstract}
\maketitle

Experimental studies of unmeasured or precisely predicted quantities
advance our knowledge of interaction mechanisms on short distances and
shed light on potential new physics.  Besides the Higgs boson couplings of
heavy particles in
the Standard Model, the
parameters of the neutrino oscillation matrix~\cite{Tanabashi:2018oca}
are poised to be measured more accurately in the coming years.
Improved precision on the neutrino mixing angles and
phase(s)~\cite{Maki:1962mu,Gribov:1968kq,Bilenky:1978nj,Bilenky:1987ty},
requires improved knowledge of neutrino-nucleus cross sections
and improved methods to determine neutrino flux~\cite{Meregaglia:2016vxf,Alvarez-Ruso:2017oui,Duyang:2019prb}.
Future neutrino experiments~\cite{Acciarri:2016crz,Abi:2020evt,Hyper-Kamiokande:2016dsw,Marshall:2019vdy} will reach sub-percent
statistical precision even for such a rare process as neutrino-electron scattering.
This process plays the role of ``standard candle" to 
constrain neutrino beam flux at near detectors.
As a proof of principle, the MINERvA Collaboration reduced uncertainties
of the NUMI beam flux normalization from $7.5\%$ to $3.9\%$~\cite{Park:2015eqa,Valencia:2019mkf}
by exploiting the neutrino-electron scattering channel.

Achieving the percent level precision
target of next generation experiments requires that one-loop radiative corrections,
and higher order corrections enhanced by large renormalization-group logarithms,
are properly treated.
Leading radiative corrections to purely leptonic processes can be readily evaluated
within the complete Standard Model.  For the simplest process of elastic
neutrino-electron scattering, the next-to-leading order electroweak
result was presented by Marciano and Sirlin in Refs.~\cite{Sirlin:1980nh,Marciano:1980pb},
exploiting the current algebra formalism and the onshell renormalization scheme
(see also Refs.~\cite{Aoki:1980ix,Bohm:1986rj,Hollik:1988ii,Denner:1991kt}
describing a direct evaluation of Feynman diagrams).  The authors of
Refs.~\cite{Sirlin:1980nh,Marciano:1980pb} also calculated the
electron energy spectrum accounting for radiation of one real photon
in the massless electron limit~\cite{Sarantakos:1982bp}.  Different
aspects of radiative corrections in elastic neutrino-electron
scattering were also discussed in
Refs.~\cite{Ram:1967zza,Salomonson:1974ys,Zhizhin:1975kv,Byers:1979af,Green:1980bd,Green:1980uc,Aoki:1981kq,Hioki:1981gi,Bardin:1983yb,Bardin:1983zm,Bardin:1985fg,Mourao:1989vb,Weber:1991kf,Buccella:1992xy,Bernabeu:1994kw,Bahcall:1995mm,Passera:2000ug,Akhmedov:2018wlf}.

However, even in leptonic processes, hadronic effects contribute a sizable
component of the radiative correction, and a dominant component of the error budget.
Computations using the complete Standard Model~\cite{Weinberg:1967tq,tHooft:1971ucy,Sirlin:1980nh,Marciano:1980pb}\footnote{References~\cite{Sirlin:1980nh,Marciano:1980pb} resort to quark-model evaluation of the hadronic contribution.}
cause electroweak scale physics and hadronic physics to be intermingled.
Four-fermion effective theory~\cite{Fermi:1934hr,Feynman:1958ty,Sudarshan:1958vf} systematically separates
contributions from electroweak and hadronic scales.   Electroweak scale physics is
computed perturbatively and is represented by four-fermion operator coefficients. 
The effective operator matrix elements provide a rigorous starting point for
nonperturbative (e.g. lattice QCD) evaluation and/or experimental analysis 
of the relevant hadronic amplitudes.    

In this work, we construct the low-energy effective field theory of neutrino-lepton and
neutrino-quark interactions suitable for predictions with
sub-percent accuracy.   Predictions for low-energy processes including  
bremsstrahlung and virtual QED corrections can
be evaluated using this theory as starting point (see e.g. Ref.~\cite{Tomalak:2019ibg} for an application
to neutrino-electron scattering).
We match the effective theory to the Standard Model (SM) at the
electroweak (EW) energy scale, evaluating all effective couplings and determining
scale-independent combinations.  We then solve renormalization
group equations (RGEs) and heavy quark threshold matching conditions,
and present four-fermion coefficients at the GeV energy scale in $\mynf=3$ and $\mynf=4$
flavor QCD.  We also present coefficients in the extreme low-energy QED limit and
determine neutrino charge radii.  

\section{Effective theory and tree level matching}

Neglecting corrections of order $\alpha m_f^2/m_W^2$ and $|q^2|/m_W^2$,
where
$m_f$ denotes a fermion mass, and $q^2$ denotes an invariant momentum transfer,
the structure of neutrino interactions with quarks and charged leptons
in the Standard Model is severely constrained.%
\footnote{
  Power corrections to four-fermion theory can be readily estimated for specific processes.
  For neutrino-electron scattering, these corrections are negligible, at the $10^{-6}$ level
  for $E_\nu \lesssim 10\,{\rm GeV}$~\cite{Tomalak:2019ibg}. 
  For neutrino-nucleon scattering, leading corrections to four-fermion theory
  scale as $|q^2|/M_W^2 \lesssim 2 m_{N} E_\nu /M_W^2$ where $m_N\approx 1\,{\rm GeV}$ is the nucleon mass
  and $E_\nu$ is the neutrino beam energy.  For $E_\nu \sim {\rm few}\,\times{\rm GeV}$, these corrections
  can amount to few-permille contributions.  However, form factors in hadronic amplitudes typically suppress 
  the region of large $|q^2|$ making these corrections further subdominant.
  }
Besides the usual 
kinetic, mass, and QED+QCD gauge coupling terms for neutrinos, charged leptons, and quarks
(including the mass mixing matrix of neutrinos), the effective Lagrangian
consists of dimension 6 four-fermion operators and neutrino-photon couplings, 
\begin{align}
    {\cal L}_{\rm eff} &=
    - \sum_{\ell,\ell^\prime}  \bar{\nu}_\ell \gamma^\mu \mathrm{P}_\mathrm{L} \nu_\ell
    \, \bar{\ell}^\prime \gamma_\mu (c_\mathrm{L}^{\nu_\ell \ell^\prime} \mathrm{P}_\mathrm{L}
    + c_\mathrm{R}^{\nu_\ell \ell^\prime} \mathrm{P}_\mathrm{R}) \ell^\prime
    -  \sum_{\ell,q}  \bar{\nu}_\ell \gamma^\mu \mathrm{P}_\mathrm{L} \nu_\ell \,
    \bar{q} \gamma_\mu (c_\mathrm{L}^{q} \mathrm{P}_\mathrm{L} + c_\mathrm{R}^{q} \mathrm{P}_\mathrm{R}) q \,
    \nl
    &\quad
    -   c \sum_{\ell \ne \ell^\prime} 
  \bar{\nu}_{\ell^\prime }\gamma^\mu \mathrm{P}_\mathrm{L} \nu_{\ell}
  \, \bar{\ell} \gamma_\mu \mathrm{P}_\mathrm{L}  \ell^\prime   - \sum_{q,\, q^\prime}  \left( c^{q q'} 
  \bar{\ell} \gamma^\mu \mathrm{P}_\mathrm{L} \nu_{\ell}
  \, \bar{q} \gamma_\mu \mathrm{P}_\mathrm{L}  q^\prime + \mathrm{h.c.}\right)
  - \frac{1}{e}\sum_\ell  c^{\nu_\ell \gamma}  \partial_\mu F^{\mu\nu} \bar{\nu}_\ell \gamma_\nu \mathrm{P}_\mathrm{L} \nu_\ell 
  \, ,  \label{effective_Lagrangian_all}
\end{align}
where $e$ denotes the positron charge.
The sums in Eq.~(\ref{effective_Lagrangian_all}) run over 3 lepton flavors ($\ell=e,\mu,\tau$) and $\mynf$ active quark flavors
($q=u,c$ and $q^\prime=d,s,b$ for $\mynf=5$).
$\mathrm{P}_\mathrm{L}= (1-\gamma_5)/{2}$ and $\mathrm{P}_\mathrm{R}=(1+\gamma_5)/{2}$
are projection operators onto left- and right-handed
states, respectively. For neutrino scattering applications, it is convenient to replace the neutrino-photon operator by an
equivalent combination of four-fermion operators in the effective theory, obtained by the field redefinition
\begin{align}
  A^\mu  \to  A^\mu + \frac{1}{e} \sum_\ell c^{\nu_\ell \gamma} \bar{\nu}_\ell \gamma^\mu \mathrm{P}_\mathrm{L} \nu_\ell  \,.
  \label{eq:field_redef}
\end{align}
Under Eq.~(\ref{eq:field_redef}), charged current coefficients $c$ and $c^{qq^\prime}$ remain unchanged while in
the neutral current sector
\begin{align}
  c^{\nu_\ell \gamma} \to 0 \,, \quad
  c_{\mathrm{L}}^{\nu_\ell \ell^\prime} \to   c_{\mathrm{L}}^{\nu_\ell \ell^\prime} + c^{\nu_\ell \gamma} \,, \quad
  c_{\mathrm{R}}^{\nu_\ell \ell^\prime} \to   c_{\mathrm{R}}^{\nu_\ell \ell^\prime} + c^{\nu_\ell \gamma} \,, \quad
  c_{\mathrm{L}}^{q} \to   c_{\mathrm{L}}^{q} -Q_q c^{\nu_\ell \gamma} \,, \quad
  c_{\mathrm{R}}^{q} \to   c_{\mathrm{R}}^{q} -Q_q c^{\nu_\ell \gamma} \,.  
  \label{eq:field_redef_couplings}
\end{align}

The four-fermion coefficients can be determined order-by-order in perturbation theory by 
matching amplitudes in the full (SM) theory and the effective theory.
In the $\overline{\rm MS}$ renormalization scheme~\cite{tHooft:1973mfk}, the lepton coefficients may be written
\begin{align}\label{eq:cldef}
  c\left(\mu \right)
  =  \frac{2 \pi \alpha \left( \mu \right)}{M^2_W(\mu) s_W^2(\mu) }  g(\mu) \,, \quad 
  c_\mathrm{L}^{\nu_\ell \ell^\prime}\left(\mu \right)
  = \left[ g_\mathrm{L}(\mu) + \delta_{\ell\ell^\prime} g(\mu) \right] \frac{c \left(\mu \right)}{g \left( \mu \right)} \,, \quad
  c^{\nu_\ell \ell^\prime}_\mathrm{R}\left(\mu \right) =  g_\mathrm{R}( \mu ) \frac{c \left(\mu \right)}{g \left( \mu \right)} \,.
\end{align}
For the quark coefficients, we have
\begin{align}\label{eq:cdef}
  c^{q q^\prime} \left(\mu \right)
  =  \frac{2 \pi \alpha \left( \mu \right)}{M^2_W(\mu) s_W^2(\mu) }  V_{q q^\prime} \left[ g(\mu) + \delta g(\mu) \right] \,, \quad 
  c_\mathrm{L}^{q}\left(\mu \right)
  = g_\mathrm{L}^{q}(\mu)  \frac{c \left(\mu \right)}{g \left( \mu \right)} \,, \quad
  c^{q}_\mathrm{R}\left(\mu \right) =  g_\mathrm{R}^{q}( \mu ) \frac{c \left(\mu \right)}{g \left( \mu \right)} \,, 
\end{align}
where for up-type quarks $q$, and down type quarks $q^\prime$, $V_{qq^\prime}$
is the corresponding element of the Cabibbo-Kobayashi-Maskawa (CKM) matrix~\cite{Cabibbo:1963yz,Kobayashi:1973fv}.
We write $M_W$ and $M_Z$ for the mass of the $W^\pm$ and $Z^0$ bosons,
related as $M_W = M_Z \cos\theta_W$, using the notation $s_W = \sin\theta_W$ and $c_W = \cos\theta_W$. 
The neutrino-photon coupling is expressed as
\ber
  c^{\nu_\ell \gamma}\left(\mu \right) =  g^{\nu_\ell \gamma}( \mu ) \frac{c \left(\mu \right)}{g \left( \mu \right)} \,.
\eer
At tree level, we have the well-known expressions~\cite{Weinberg:1967tq,tHooft:1971ucy}
for lepton couplings,
\begin{align}\label{eq:leptongtree}
  g = 1 \,, \qquad
  g_\mathrm{L}^{\mathrm{0}} = s_W^2 - \frac12 \,, \qquad
  g_\mathrm{R}^{\mathrm{0}} = s_W^2 \,, 
\end{align}
and for quark couplings, 
\begin{align}
  \delta g=0 \,, \qquad
  g_\mathrm{L}^{q\mathrm{0}} = T^3_q - Q_q s_W^2 \,, \qquad
  g_\mathrm{R}^{q\mathrm{0}} = -Q_q s_W^2 \,. 
\end{align}
The neutrino-photon coupling vanishes at tree level, $g^{\nu_\ell \gamma}=0$ and $c^{\nu_\ell \gamma} =  \order(\alpha \mathrm{G_F}) $\,. 
Here $Q_q$ is the quark electric charge in units of the positron charge ($Q_u=2/3$, $Q_d=-1/3$), $\mathrm{T}^3_q$
is the quark isospin ($\mathrm{T}^3_u = +1/2$, $\mathrm{T}^3_d = -1/2$),
and $\alpha(\mu)$ is the electromagnetic coupling constant.

\section{One-loop electroweak matching}

We perform the matching onto the effective theory (\ref{effective_Lagrangian_all}) 
by first integrating out heavy vector and scalar bosons, $W^\pm$, $Z^0$ and $H$,
and the top quark, $t$, in the Standard Model.  
Before the field redefinition (\ref{eq:field_redef}), the neutrino-photon
coupling is determined by computing the neutrino scattering process
$\nu_\ell(p) \to \nu_\ell(p^\prime)$ in a background electromagnetic field.\footnote{The coefficient $g^{\nu_\ell \gamma}( \mu ) $ is determined by $\gamma Z$-mixing diagrams, penguin-type diagrams with $W$, and the closed top-loop contribution on the Standard Model side subtracting appropriate contributions in the effective theory.}
After $\overline{\rm MS}$ renormalization, the one-loop contribution in Feynman-t' Hooft gauge is
\begin{align} \label{eq:photong}
g^{\nu_\ell\gamma}( \mu )  & =
- \frac{\alpha}{8 \pi} \left(  9  - 10 s_W^2 \right) \ln \frac{\mu^2}{M^2_W}
- \frac{\alpha }{ 12\pi} \left(  5 - 2  s_W^2  \right)
+ \frac{\alpha}{18\pi} \left( 3 - 8 s_W^2 \right)  \ln \frac{\mu^2}{m_t^2}  
\,.
\end{align}
After the field redefinition (\ref{eq:field_redef}), this coefficient vanishes $g^{\nu_\ell\gamma} \to 0$ 
and all other couplings change according to Eq.~(\ref{eq:field_redef_couplings}).
Reproducing one-loop EW radiative
corrections in the Standard
Model~\cite{Sirlin:1980nh,Marciano:1980pb,Bohm:1986rj,Hollik:1988ii,Denner:1991kt}
in Feynman-'t Hooft gauge with massless leptons and quarks, besides
the top, and subtracting the corresponding diagrams in the effective
theory, we obtain the following reduced couplings for neutrino-lepton interactions after the field redefinition~(\ref{eq:field_redef_couplings}): 
\begin{align} \label{eq:leptong}
g_\mathrm{L} \left( \mu \right) &= g_\mathrm{L}^{\mathrm{0}} \Bigg( 1
-  \frac{\alpha}{16\pi c_W^2 s_W^2} \bigg[  6 \mytau \ln \frac{\mu^2}{m^2_t} - \ln \frac{\mu^2}{M^2_Z}
  +  \frac{3}{1 - \myxi} \left(\myxi \ln \frac{\mu^2}{M^2_H}
  -  \ln \frac{\mu^2}{M^2_Z} \right) + \frac{\myxi -7} {2}
  + 6 g_\mathrm{L}^{\mathrm{0}} \bigg]
\nl
&\quad 
+ \frac{ \alpha}{2 \pi}  \ln \frac{\mu^2}{M^2_W}  \Bigg)
- \frac{\alpha}{4\pi} \frac{  c_W^2}{s_W^2}  \ln \frac{\mu^2}{M^2_W}
- \frac{ \alpha}{2 \pi s_W^2}
+ {g}^{\nu_\ell \gamma}
\,,
\nl
g_\mathrm{R} \left( \mu \right) &=  g_\mathrm{R}^{\mathrm{0}} \Bigg( 1 -  \frac{\alpha}{16 \pi c_W^2 s_W^2} \bigg[  6 \mytau \ln \frac{\mu^2}{m^2_t} - \ln \frac{\mu^2}{M^2_Z} +  \frac{3}{1 - \myxi} \left(\myxi \ln \frac{\mu^2}{M^2_H} -  \ln \frac{\mu^2}{M^2_Z} \right)   + \frac{\myxi-7}{2} - 6 g_\mathrm{R}^{\mathrm{0}} \bigg] \nl
&\quad
+ \frac{ \alpha}{2 \pi}  \ln \frac{\mu^2}{M^2_W} \Bigg)
+ {g}^{\nu_\ell \gamma}
\,,
  \nl
  g \left( \mu \right) &= 1 -  \frac{\alpha}{16 \pi c_W^2 s_W^2 } \bigg[ 6 \mytau \left( \ln \frac{\mu^2}{m^2_t} +\frac{1}{2} \right)  - c^2_W  \ln \frac{\mu^2}{M^2_W} +  \frac{3 c_W^2}{c_W^2 - \myxi} \left( \myxi \ln \frac{\mu^2}{M^2_H} - c_W^2  \ln \frac{\mu^2}{M^2_W}  \right)
    \nl
    &\quad
    + \frac{\myxi - 7 }{2}  \bigg] 
  + \frac{\alpha}{4\pi s_W^2 c_W^2} \left( 1 +  c_W^2 \right)\ln \frac{\mu^2}{M^2_W}
  + \frac{\alpha}{16 \pi s_W^4 c_W^2} \left( 7 s_W^2 - 3  \right)\ln \frac{M^2_W}{M^2_Z} + \frac{7\alpha}{16\pi s_W^2}  ,
 \end{align}
where $\mytau = m^2_t/M^2_Z$ and $ \myxi= M^2_H/M^2_Z$ with masses of the top quark
$m_t$ and the Higgs boson $M_H$.  Tree level couplings 
$g_\mathrm{L,R}^{\mathrm{0}}$ have been specified above and depend on 
renormalization scale $\mu$ through $s_W^2$.

To perform the matching, we have enforced the vanishing of the Higgs tadpole
(the amputated and renormalized Higgs field one-point function)
as a renormalization condition, but use ${\overline{\rm MS}}$
renormalization for non-tadpole counterterms.
This hybrid renormalization scheme leads to compact, but gauge-dependent, expressions
for the reduced couplings (\ref{eq:leptong}) and for $\overline{\mathrm{MS}}$ masses.
The prefactor, $\alpha M_W^{-2} s_W^{-2}$, in Eq.~(\ref{eq:cldef}), 
and the tree level reduced couplings in (\ref{eq:leptongtree}) must be evaluated in this scheme
in order to employ the expressions (\ref{eq:leptong}).
Here $\alpha(\mu)$, as well as $M_W$, $M_Z$ and $s_W$,
refer to the full SM particle content.   
We have verified that the complete effective couplings (\ref{eq:cdef}) are gauge independent
and have obtained the same one-loop matching onto the effective Lagrangian of
Eq.~(\ref{effective_Lagrangian_all}) in an arbitrary $\mathrm{R}_\xi$ gauge.%
\footnote{See
Refs.~\cite{Degrassi:1989ip,Degrassi:1992ff,Degrassi:1992ue} for the
discussion of one-loop EW corrections in $\mathrm{R}_\xi$ gauges.}
We have also obtained the one-loop coefficients using on-shell renormalization,
first expressing the effective couplings (\ref{eq:cdef}) in terms of on-shell
(pole) $M_W$, $M_Z$, and low-energy (Thomson limit) $\alpha$, and then
expressing these on-shell quantities in terms of $M_W(\mu)$, $M_Z(\mu)$
and $\alpha(\mu)$.
In performing the matching, we have regulated infrared divergences in both
full and effective theories with small photon and fermion masses.
The matching could be performed in the exact massless limit but would require
consideration of new operator structures in the effective theory
to account for Fierz rearrangements in $d\ne 4$.

Similar to Eq.~(\ref{eq:leptong}), we can determine neutrino-quark couplings as
\begin{align} \label{eq:quarkg}
  g^{q}_\mathrm{L} \left( \mu \right)
  &=
  g^{q \mathrm{0}}_\mathrm{L}  \Bigg( 1 -  \frac{\alpha}{16\pi s_W^2 c_W^2}    \bigg[  6 \mytau \ln \frac{\mu^2}{m^2_t} - \ln \frac{\mu^2}{M^2_Z} +  \frac{3}{1 - \myxi} \left(\myxi \ln \frac{\mu^2}{M^2_H} -  \ln \frac{\mu^2}{M^2_Z} \right) + \frac{\myxi -7} {2} +
    6 g^{q\mathrm{0}}_\mathrm{L}\bigg]
  \nl
  &\quad 
    + \frac{ \alpha}{2 \pi}  \ln \frac{\mu^2}{M^2_W}  \Bigg)
  + \frac{ \alpha}{16\pi s_W^2} \left( 10 \mathrm{T}^3_q -3 \right)
  + \frac{\alpha}{2\pi} \frac{c_W^2 }{s_W^2} \mathrm{T}^3_q \ln \frac{\mu^2}{M^2_W}
 - Q_q {g}^{\nu_\ell \gamma}
  \nl
  &\quad
  +  \frac{\alpha}{16 \pi s_W^2} \frac{\omega_t}{\omega_t -1} \left( \frac12 - \mathrm{T}^3_q  \right)  |V_{tq}|^2
  \left(2 + \omega_t + 3 \frac{\omega_t - 2 }{\omega_t - 1}  \ln \omega_t \right) \,,
  \nl
  g^{q}_\mathrm{R} \left( \mu \right)
  &=
  g^0_\mathrm{R} \Bigg( 1 -  \frac{\alpha}{16 \pi s_W^2 c_W^2}   \bigg[  6 \mytau \ln \frac{\mu^2}{m^2_t} - \ln \frac{\mu^2}{M^2_Z} +  \frac{3}{1 - \myxi} \left(\myxi \ln \frac{\mu^2}{M^2_H} -  \ln \frac{\mu^2}{M^2_Z} \right)
    + \frac{\myxi-7}{2} - 6 g^0_\mathrm{R} \bigg]
\nl
&\quad
+ \frac{ \alpha}{2 \pi}  \ln \frac{\mu^2}{M^2_W}\Bigg)
- Q_q {g}^{\nu_\ell \gamma} \,,
  \nl
\delta g(\mu) &= - \frac{\alpha}{2 \pi}  \ln \frac{\mu^2}{M^2_Z} - \left( 1 - a \right)\frac{\alpha}{6 \pi} \,,
\end{align}
where $\omega_t = m_t^2/M_W^2$.
Here $a$ stands for the scheme parameter defining the behavior of $\gamma_5$ in $d\ne 4$.  
It appears in the reduction of Dirac matrices resulting from two-boson exchange diagrams, e.g. 
\ber
\gamma^\alpha \gamma^\beta \gamma^\mu \mathrm{P}_\mathrm{L} \otimes \gamma_\mu \gamma_\beta \gamma_\alpha \mathrm{P}_\mathrm{L}
= 4 \left[ 1 + a \left( 4-d\right) \right] \gamma^\mu \mathrm{P}_\mathrm{L} \otimes \gamma_\mu \mathrm{P}_\mathrm{L} + \mathrm{E} \left( a \right), \label{evanescent}
 \eer
 where $\mathrm{E}$ is an evanescent operator with vanishing matrix element in $d=4$.
 We choose $a = -1$ in the following, appropriate for anticommuting $\gamma_5$ in the  basis
 ${\gamma^\mu\otimes \gamma_\mu,~\gamma^\mu \otimes \gamma_\mu \gamma_5,~\gamma_5 \gamma^\mu \otimes \gamma_\mu,~\gamma_5 \gamma^\mu \otimes \gamma_\mu \gamma_5 }$~\cite{Buras:1989xd,Dugan:1990df,Herrlich:1994kh}.
 We remark that the difference between left-handed couplings to $s$ and $d$ quarks in Eq.~(\ref{eq:quarkg}) 
 is negligibly small, using $|V_{tb}|^2 = 1 - |V_{td}|^2 - |V_{ts}|^2 \approx 1$. 
 However, the top quark contributes significantly to the left-handed coupling of the $b$ quark.

The starting point for our renormalization analysis below is $\mynf=5$ flavor QCD,
with four-fermion coefficients determined at $\mu=M_Z$\footnote{\label{note1}For definiteness, we employ the mass of $Z$ boson,
  $M_Z^{\rm PDG}$ from Eq.~(\ref{eq:heavymassinput}), as the scale $\mu$ for
  all $\overline{\mathrm{MS}}$ quantities at $\mu = M_Z$.} through $\order(\alpha \GF)$
in Eqs.~(\ref{eq:leptong}) and (\ref{eq:quarkg}).
For lepton coefficients (\ref{eq:leptong}), we also include complete corrections
through
$\order(\alpha \alpha_s \GF)$~\cite{Kniehl:1989yc,Degrassi:1990tu,Fanchiotti:1992tu,Djouadi:1993ss,Avdeev:1994db,Chetyrkin:1995ix,Erler:1998sy},
arising from gluons attached to closed quark loops.  
Details about this correction, and the determination of appropriate $\overline{\rm MS}$ masses,
are discussed in the Appendix~\ref{app:MSbar_from_observables}.
Uncomputed corrections of $\order(\alpha \alpha_s \GF)$ to quark coefficients are left to
future work; these corrections are small compared to hadronic uncertainties in current
neutrino scattering analyses. 
Note that in Eqs.~(\ref{eq:leptong}) and (\ref{eq:quarkg}) we have performed the
field redefinition (\ref{eq:field_redef}) and we maintain the
condition $c^{\nu_\ell\gamma}=0$ for renormalized coefficients.

\section{Standard Model inputs \label{sec:inputs}}

\begin{figure}[t]
\begin{center}
\vspace{-0.25cm}	
\hspace{0.7cm}	\includegraphics[scale=1.]{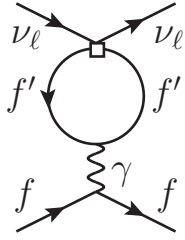}\hspace{0.cm}
\end{center}
\caption{Leading contribution to running of couplings in effective theory.\label{one_loop_ET}
}
\end{figure}

Having computed expressions for the four-fermion operator coefficients, 
let us define the numerical inputs to these expressions.  We begin by
isolating coefficient combinations that are independent of renormalization
scale, and combinations that vanish up to neglected $\order(\alpha m_f^2/m_W^2)$ corrections.  

As a consequence of Ward-Takahashi identities~\cite{Ward:1950xp,Takahashi:1957xn}
and the partial conservation of vector and axial-vector currents in
 QED+QCD gauge theory, the evolution equation for couplings in
the neutral current sector is governed by 
vacuum polarization diagrams as in Fig.~\ref{one_loop_ET}, 
\begin{align}
  \frac{\mathrm{d}}{\mathrm{d} \ln \mu^2}{ c^{f}_\mathrm{S} \over Q_f }= {\alpha \over 6\pi}  \bigg[
    \sum_{f^\prime,S^\prime} N_{c,f^\prime} Q_{f^\prime} c^{f^\prime}_{S^\prime}
    \bigg]  + \dots \,,
\end{align}
where $f$ denotes an active lepton or quark flavor in the theory,
$N_{c,f}$ is the number of color degrees of freedom ($N_{c,f}=1$ for leptons,
$N_{c,f}=3$ for quarks), and $S=\mathrm{L},\mathrm{R}$. 
In particular, the running of $c^{f}_S/Q_f$ does
not depend on $f$ or $S$.  This property of the anomalous dimension matrix
holds also at higher orders so that%
\footnote{Diagrams as in Fig.~\ref{one_loop_ET} with
  the exchanged photon replaced by a gluon vanish by color symmetry.
   Diagrams with two photons exchanged between the loop in Fig.~\ref{one_loop_ET} 
   and exterior fermion legs
vanish in the sum of direct and crossed contributions according to
Furry's theorem.
The sum of three-photon exchange contributions is UV
finite~\cite{Aldins:1970id,Lautrup:1977tc} and does not affect the
running.
Similar arguments apply to two- and three-gluon exchange
diagrams.  While not phenomenologically important,
it may be formally interesting to investigate this
structure to higher orders in QED+QCD couplings.
}
\begin{align}
{c_S^f(\mu) \over Q_f} - {c_{S^\prime}^{f^\prime}(\mu) \over Q_{f^\prime}} = {\rm constant} \,.
\end{align}
All couplings in the neutral current sector may thus be written as
linear combinations of scale-invariant quantities and a single scale-dependent
coupling.
Neglecting fermion mass corrections of order $\alpha m_f^2/m_Z^2$, we also have (at $\mu=M_Z$):
$c_{\rL,\rR}^{\nu_\ell \ell^\prime} = c_{\rL,\rR}^{\nu_e e}$  ($\ell=\ell^\prime = e,\mu,\tau$),
$c_{\rL,\rR}^{\nu_\ell \ell^\prime} = c_{\rL,\rR}^{\nu_\mu e}$ ($\ell \ne \ell^\prime$),
$c_{\rR}^{\nu_\mu e} = c_{\rR}^{\nu_e e}$, 
$c_{\rR}^{b}= c_{\rR}^{d}$,
$c_{\rL,\rR}^{c}=c_{\rL,\rR}^{u}$,
$c_{\rL,\rR}^{s}=c_{\rL,\rR}^{d}$. 

In the charged current leptonic sector described by coefficient $c(\mu)$,
the operator anomalous dimension vanishes,
and we may define the scale-independent Fermi constant~\cite{Antonelli:1980zt,Arason:1991ic}%
\ber
c \left(\mu \right) = c \equiv 2 \sqrt{2} \mathrm{G}_\mathrm{F}. \label{Fermi_running}
\eer
In the charged current semileptonic sector, the presence of three external charged fermion
lines results in scale dependence of $c^{qq^\prime}(\mu)$, cf. Eq.~(\ref{eq:RG_CC}) below. 
The remaining independent coefficients are the scale-dependent quantity
$c_{\mathrm{R}}(\mu) \equiv c_{\mathrm{R}}^{\nu_\mu e}(\mu)$, and
the scale-independent combinations
\begin{align} \label{scale_invariant}
  c^{\nu_e e}_\mathrm{L}  - c^{\nu_\mu e}_\mathrm{L}  
  &= 2 \sqrt{2} \mathrm{G}_\mathrm{F}, \nl
  3 c_\mathrm{L}^u + 2 c_\mathrm{L}^{\nu_\mu e}
  &= \,\,\, \sqrt{2}  \mathrm{G}_{u}\,,\nl
  - 3 c_\mathrm{L}^d + c_\mathrm{L}^{\nu_\mu e}
  &= 2\sqrt{2}  \mathrm{G}_d\,,\nl
  c_\mathrm{L}^{\nu_\mu e} - c_\mathrm{R}
  &= - \sqrt{2} \mathrm{\tilde{G}}_e \,, \nl
  c_\mathrm{L}^{u } - c^u_\mathrm{R}
  &= \,\,\,\,\sqrt{2} \mathrm{\tilde{G}}_u \,, \nl
  c_\mathrm{L}^{d} - c^d_\mathrm{R} 
  &= -\sqrt{2} \mathrm{\tilde{G}}_d \,, \nl
  c_\mathrm{L}^{b} - c^b_\mathrm{R}
  &= -\sqrt{2} \mathrm{\tilde{G}}_b \,.
\end{align}
The quantities 
$\mathrm{G}_q$ ($q=u,d$) are differences of left-handed couplings $c_\mathrm{L}^q$
and $c_\mathrm{L}^{\nu_\mu e}$,
while tilded quantities
$\tilde{\mathrm{G}}_f$ ($f=e,u,d,b$) are differences of $c_\mathrm{L}^f$ and $c_\mathrm{R}^f$. 

Having reduced the scale dependence to $c_\mathrm{R}(\mu)$ in the neutral current sector
and $c^{qq^\prime}(\mu)$ in the charged current sector, we proceed to specify
numerical inputs to the intial conditions of RGE. 
For numerical evaluation, we employ high order running and threshold
matching corrections for $\overline{\rm MS}$ QCD and QED couplings%
\footnote{\label{note2}Evolution of $\alpha_s(\mu)$ is computed with five-loop QCD running and four-loop threshold matching corrections,
  ignoring QED. 
  Evolution of $\alpha(\mu)$ is computed with two-loop QED and $\order(\alpha_s^3)$ corrections 
  to running, and one-loop QED and $\order(\alpha_s^2)$ corrections to threshold matching.
}
with input values~\cite{Tanabashi:2018oca}
\begin{align}\label{eq:alphainput}
  \alpha_s^{(5)}(\mu=M_Z) = 0.1187(16) \,, \quad
  \alpha^{(5)}(\mu=M_Z)^{-1} = 127.955(10) \,.
\end{align}
The $\overline{\rm MS}$ quantities $\alpha^{(5)}$ and $\alpha_s^{(5)}$ in Eq.~(\ref{eq:alphainput}) have been
defined in the theory containing $W^\pm$, $Z^0$ and $H$, i.e., the complete SM particle content except the top quark.
For values at lower scales and in the theory with top quark, see Appendix~\ref{app:alpha}.
We use the Fermi constant determined from muon decay,
\begin{align}\label{eq:GFinput}
\GF = 1.1663787(6)\times 10^{-5}\,{\rm GeV}^{-2} \,,
\end{align}
and the following mass parameters~\cite{Tanabashi:2018oca}
\begin{align}\label{eq:heavymassinput}
  M_Z^{\rm PDG} = 91.1876(21)~\mathrm{GeV} \,,  \qquad
  M_H = 125.10 (16)~\mathrm{GeV} \,, \qquad
  m_t^{\rm PDG} = 172.9(4)~\mathrm{GeV} \,.
\end{align}
The quantity $M_Z^{\rm PDG}$ represents a fit parameter in $Z^0$ lineshape analysis; its relation to the
corresponding pole mass and $\overline{\rm MS}$ mass is discussed in the Appendix~\ref{app:MSbar_from_observables}. 
The quantity $m_t^{\rm PDG}$ represents the top quark pole mass.  The translation to the corresponding
$\overline{\rm MS}$ mass, relevant for $\order(\alpha \alpha_s)$ corrections to four-fermion coefficients, is
discussed in the Appendix~\ref{app:MSbar_from_observables}. 
The  quantities  (\ref{eq:alphainput}), (\ref{eq:GFinput}), and (\ref{eq:heavymassinput}),
together with CKM elements for semileptonic charged current processes, fully determine the
electroweak scale matching coefficients.

\begin{table*}
\centering
\caption{Neutrino-lepton four-fermion coefficients.
  $\GF$ and $\mathrm{\tilde{G}_e}$ are in units $10^{-5}~\mathrm{GeV^{-2}}$.  
  The second and third lines of the table provide our evaluation using
  inputs ($M_W$, $M_Z$, $\alpha^{(5)}$) and ($\GF$, $M_Z$, $\alpha^{(5)}$)
  respectively. Uncertainty due to Standard Model input
  parameters is denoted with index $\mathrm{i}$.  Perturbative uncertainty is denoted with index $\mathrm{p}$.
}
\label{results_couplings}
\begin{minipage}{\linewidth}  
\centering
\begin{tabular}{|l|c|c|c|c|c|c|c|}
  \hline          
  inputs &$ \mathrm{G}_\mathrm{F}$&$ \mathrm{\tilde{G}}_{e}$ & $\dfrac{c_\mathrm{R}(M_Z)}{2\sqrt{2}\GF}$ 
 & $s^2_W(M_Z)$
\\
\hline
tree level  &$1.12508 (85)_\mathrm{i}$&$1.12508 (85)_\mathrm{i}$&$0.22301(23)_\mathrm{i}$ & $0.22301(23)_\mathrm{i}$\\
$M_W,M_Z,\alpha^{(5)}$ &$1.16713 (83)_\mathrm{i}(12)_\mathrm{p}$&$1.18161(85)_\mathrm{i}(33)_\mathrm{p}$&$0.22998(7)_\mathrm{i}$ & $0.23123(23)_\mathrm{i}$\\
$\mathrm{G}_\mathrm{F},M_Z,\alpha^{(5)}$  &$1.1663787(6)_\mathrm{i}$& $1.18083(4)_\mathrm{i}(21)_\mathrm{p}$  & $0.23004(3)_\mathrm{i}$ & $0.23144(3)_\mathrm{i}$\\
\hline 
\end{tabular}
\end{minipage}
\end{table*}

Leptonic and semileptonic four-fermion coefficients at $\mu=M_Z$ are
displayed in Tables~\ref{results_couplings} and \ref{results_couplings_quark},
respectively. 
For reference, we also display the determination of
$s^2_W(\mu=M_Z)$ in the tables.%
\footnote{Note that we employ the $\overline{\rm MS}$ scheme for $s^2_W$, in contrast to
  the quantity $\hat{s}^2_Z$ of Ref.~\cite{Tanabashi:2018oca} which includes also a finite
  subtraction~\cite{Fanchiotti:1992tu}.  Our convention corresponds to the
  quantity $\hat{s}^2_{\rm ND}$ of Ref.~\cite{Tanabashi:2018oca}.
}
\footnote{Choosing the right-handed coefficient $c_\mathrm{R} \left( \mu \right)$
  as the relevant scale-dependent coupling avoids difficulties in defining
  $s^2_W$ at low energies~\cite{Erler:2004in,Erler:2017knj,Erler:2013xha}.
}
To illustrate the overall size of electroweak corrections at $\mu=M_Z$,
the first line of the tables shows the tree level evaluation of these quantities.
For this purpose, we extract the weak mixing angle from pole masses $c_W = M_W / M_Z$ and take
the electromagnetic coupling constant $\alpha$ in the Thomson limit~\cite{Tanabashi:2018oca},
$\alpha_0^{-1} = 137.035999139(31)$.
Our final results for four-fermion coefficients employ the inputs
(\ref{eq:alphainput}), (\ref{eq:GFinput}), and (\ref{eq:heavymassinput}),
and are given by the third line of the tables.
For comparison, the second line of the tables presents coefficents determined without the muon lifetime
constraint, employing $M_W^\mathrm{PDG}=80.379(12)\,{\rm GeV}$ in place of $\GF$.

The uncertainties displayed in Tables~\ref{results_couplings} and \ref{results_couplings_quark}
correspond to Standard Model input parameters (denoted by index ``i"),
and to uncalculated higher-order perturbative corrections (denoted by index ``p"). 
The leading unaccounted corrections appear at 
$\order\left(\alpha \alpha_s^2, \, \alpha^2
\right)$
for leptonic coefficients and at
$\order\left(\alpha \alpha_s \right)$ for quark coefficients.%
\footnote{For a discussion of higher-order corrections,
  see Refs.~\cite{Kataev:1992dg,Barbieri:1992dq,Jegerlehner:2001fb,Jegerlehner:2002em,Degrassi:2014sxa,Martin:2015rea,Martin:2015lxa}.}
The perturbative uncertainty, at the level  $10^{-4}$, is estimated 
by varying the matching scale in the range $M_Z^2/2 \le \mu^2 \le 2 M_Z^2$
for scale-invariant quantities.
Note that the scale dependence is larger for $\mathrm{\tilde{G}}_b$ due to the
large CKM matrix element $V_{tb}$.
This uncertainty is subdominant to SM parameter input uncertainty when
using EW inputs, represented by the second line of Tables~\ref{results_couplings}
and~\ref{results_couplings_quark}.
With the $\GF$ constraint, represented by the third line of Tables~\ref{results_couplings}
and~\ref{results_couplings_quark}, perturbative uncertainty dominates but
is well below current hadronic uncertainties for neutrino scattering applications.

As an immediate by-product of our analysis, we may compare the electroweak scale
determination of $\GF$ to the low-scale muon lifetime measurement. 
Such an analysis was discussed in detail in Refs.~\cite{vanRitbergen:1999fi,Marciano:1999ih} with last
update in Ref.~\cite{Marciano:2011zz}. The precision of EW parameters
at high energies~\cite{Tanabashi:2018oca} and of
muon lifetime measurements~\cite{Barczyk:2007hp,Casella:2013bla,Webber:2010zf,Tishchenko:2012ie}
have significantly increased after the publication of
Refs.~\cite{vanRitbergen:1999fi,Marciano:1999ih,Marciano:2011zz}.
The comparison of $\mathrm{G}_\mathrm{F} = 1.16713(83)(12)\,\times 10^{-5}\,{\rm GeV}^2$
at $\mu=M_Z$ versus $\mathrm{G}_\mathrm{F} = 1.1663787(6)\,\times 10^{-5}\,{\rm GeV}^2$
at $\mu=m_\mu$ shows only $0.9\sigma$ tension with $6\times 10^{-4}$ relative difference.

\begin{table*}
\centering
\caption{Same as Table~\ref{results_couplings} but for neutrino-quark four-fermion coefficients, in units $10^{-5}~\mathrm{GeV^{-2}}$.
}
\label{results_couplings_quark}
\begin{minipage}{\linewidth}  
\centering
\begin{tabular}{|l|c|c|c|c|c|c|}
  \hline
         {inputs} & $\mathrm{\tilde{G}}_u$ &$\mathrm{\tilde{G}}_d$ &$\mathrm{\tilde{G}}_b$\\
\hline
tree level &$1.12508 (85)_\mathrm{i}$&$1.12508 (85)_\mathrm{i}$&$1.12508 (85)_\mathrm{i}$ \\
$M_W,M_Z,\alpha^{(5)}$&$1.16917(83)_\mathrm{i}(32)_\mathrm{p}$&$1.18231(85)_\mathrm{i}(33)_\mathrm{p}$&$1.16325(83)_\mathrm{i}(80)_\mathrm{p}$\\
$\mathrm{G}_\mathrm{F},M_Z,\alpha^{(5)}$&$1.16841(4)_\mathrm{i}(20)_\mathrm{p}$&$1.18154(4)_\mathrm{i}(21)_\mathrm{p}$&$1.16250(2)_\mathrm{i}(69)_\mathrm{p}$ \\  
\hline
\hline        
inputs &$\mathrm{G}_u$ &$\mathrm{G}_d$ &$\dfrac{c^{qq^\prime}(M_Z)}{2\sqrt{2} V_{qq^\prime}}$\\
\hline
tree level &$1.12508 (85)_\mathrm{i}$&$1.12508 (85)_\mathrm{i}$&$1.12508 (85)_\mathrm{i}$ \\
$M_W,M_Z,\alpha^{(5)}$&$1.14642(78)_\mathrm{i}(33)_\mathrm{p}$&$1.18288(85)_\mathrm{i}(33)_\mathrm{p}$&$1.16622 \left(83  \right)_\mathrm{i}$\\
$\mathrm{G}_\mathrm{F},M_Z,\alpha^{(5)}$&$1.14570(4)_\mathrm{i}(22)_\mathrm{p}$&$1.18211(4)_\mathrm{i}(21)_\mathrm{p}$&$1.165468 \left(4\right)_\mathrm{i}$ \\  
\hline
\end{tabular}
\end{minipage}
\end{table*}

\section{Running}

The RGE in the neutral current sector mixes all quark and lepton operators involving a given neutrino flavor: 
\begin{align}\label{eq:RG_NC}
  \frac{\mathrm{d}}{\mathrm{d} \ln \mu^2}{ c^{f}_\mathrm{S} \over Q_f }
  = \frac{1}{2} \sum_{f^\prime,S^\prime}  \beta_{f^\prime}  {c_{S^\prime}^{f^\prime}\over Q_{f^\prime}} \,.
\end{align}
Here $\beta_f$ denotes the contribution of fermion $f$ to the QED beta function:
\begin{align}
  \frac{\mathrm{d} \ln \alpha  }{\mathrm{d} \ln \mu^2} = \sum_{f} \beta_f   \,, \quad
\end{align}
where for leptons and quarks we have 
\ber
\beta_\ell = \frac{\alpha}{\pi} \left( \frac{1}{3}  + \frac{\alpha}{4 \pi} \right) \,, \qquad
\beta_{q} = Q_q^2 \mathrm{N_c} \frac{\alpha}{\pi}
\left[ \frac{1}{3} + \frac{\alpha_s}{3 \pi} + \left( \frac{125}{48} - \frac{11 \mynf}{72}  \right) \frac{\alpha_s^2}{3 \pi^2} + \frac{\alpha}{4 \pi} Q^2_q \right],
\eer
with $\mathrm{N_c}=3$.
The  $\order(\alpha_s^3)$ contribution does not change results of this work within significant digits.
In the charged current sector, the coefficient $c$ is scale independent while
the anomalous dimension of $c^{q q'}$ is not zero~\cite{Sirlin:1977sv,Sirlin:1981ie,Marciano:1985pd,Marciano:1993sh,Erler:2002mv,Altarelli:1980fi,Buras:1989xd}%
\footnote{The contribution of order $\alpha \alpha_s$ to
  anomalous dimension in Eq.~(\ref{eq:RG_CC}) is based on results of
  Ref.~\cite{Buras:1989xd} obtained within $a=-1$ scheme in Feynman-'t
  Hooft gauge. However, it is expected to be gauge
  independent~\cite{Caswell:1974cj}.}
\begin{align}\label{eq:RG_CC}
       \frac{\mathrm{d} \ln c^{q q'}  }{\mathrm{d} \ln \mu^2} =-\frac{\alpha}{2\pi} \left( 1 - \frac{\alpha_s}{4\pi}  \right) \,. 
\end{align}
Accounting for the scale-dependence of $\alpha$ and $\overline{\rm MS}$
vector-boson masses $M_W$ and $M_Z$, 
it is readily verified that the $\mu$-dependence of the one-loop matching coefficients, 
(\ref{eq:cldef}) and (\ref{eq:cdef}),
obeys the evolution equations derived from the UV structure of effective theory operators,
Eqs.~(\ref{eq:RG_NC}) and (\ref{eq:RG_CC}). 

We solve the RGE between particle thresholds
numerically for the coefficients 
$c_\mathrm{R} \left( \mu \right)$, $c^{q q'} \left(\mu \right)$, using high
order running of $\alpha \left( \mu \right)$ and $\alpha_s \left( \mu \right)$.\footnoteref{note2}
Integrating out heavy $b$ and $c$ quarks, we account for small threshold
matching effects through $\order(\alpha \alpha_s)$. 
We use the following values for bottom and charm quark masses~\cite{Tanabashi:2018oca}, 
\begin{align}\label{eq:quarkmassinput}
  m_b(m_b) = 4.18(3)~\mathrm{GeV} \,, \qquad
  m_c(m_c) = 1.28(2)~\mathrm{GeV} \,. 
\end{align}

\begin{table*}[t]
\centering
\caption{Effective couplings in the Fermi theory of neutrino-fermion
  scattering with $\mynf$ quark flavors, at different renormalization scales,
  in units $10^{-5}~\mathrm{GeV^{-2}}$.
  $\tau$ is not present in the theory described by last
  three rows.
  The final row gives couplings to $\nu_\tau$.
}
\label{results_couplings_running}
\begin{minipage}{\linewidth}
\footnotesize
\centering
\begin{tabular}{|l|c|c|c|c|c|}
\hline          
$\mu$ & $\mynf$ & $ c^{\nu_\ell \ell'}_\mathrm{L},~\ell = \ell'$ & $ -c^{\nu_\ell \ell'}_\mathrm{L},~\ell \neq \ell'$ & $ c_\mathrm{R}$ & $c^{q q'} / V_{q q'}$ \\
\hline
$M_Z$ & $5$ & $2.38798 (32)$ &$0.91103(32)$ &$0.75891(60)$  & $3.29644(1)$  \\
\hline
$m_b$ & $4$ & $2.39676(33)$ &$0.90226(32)$ &$0.76769(60)$ & $3.32110(6)$ \\
\hline
$2~\mathrm{GeV}$ & $4$ &  $2.39818(33)$ &$0.90084(32)$ &$0.76911(60)$  & $3.32685(8)$ \\
\hline
$m_\tau$ & $4$ &  $2.39841(33)$ &$0.90060(33)$ &$0.76935(60)$  & $3.32776(8)$ \\
\hline
$m_c$ & $3$ &  $2.39912(35)$ &$0.89989 (35)$ &$0.77006 (61)$  & $3.33029(9) $ \\
&
&  --- &$0.89904 (35)$ &$0.77092 (61)$ & ---  \\
\hline
\hline
$\mu$ & $\mynf$ &  $ c^{u}_\mathrm{L}$
& $ -c^{u}_\mathrm{R}$ & $ -c^{d}_\mathrm{L}$ & $ c^{d}_\mathrm{R}$   \\
\hline
$M_Z$ & $5$ & $1.14745(13)$ &$0.50494(38)$ &$1.41818(12)$ &$0.25277(20)$  \\
\hline
$m_b$ & $4$ & $1.14160 (13)$ &$0.51079 (38)$ &$1.41525 (12)$ &$0.25570 (20)$  \\
\hline
$2~\mathrm{GeV}$ & $4$ & $1.14065 (13)$ &$0.51173 (38)$ &$1.41478 (12)$ &$0.25617 (20)$  \\
\hline
$m_\tau$ & $4$ &$1.14049(14)$ &$0.51189(38)$ &$1.41470(12)$ &$0.25625(20)$ \\
\hline
$m_c$ & $3$ & $1.14002 (16)$ &$0.51236 (39)$ &$1.41447(13)$ &$0.25648 (20)$  \\
&
& $1.13945 (16)$ & $0.51294 (39)$ &$1.41418 (13)$ &$0.25677 (20)$ \\
\hline
\end{tabular}
\end{minipage}
\end{table*}

The running from EW scale to GeV energies yields
the effective couplings in Table~\ref{results_couplings_running}. 
We have employed the weak scale couplings from the last row of each of 
Tables~\ref{results_couplings} and \ref{results_couplings_quark}.
The uncertainty from higher-order perturbative corrections
is estimated by varying matching scales
$\mu_Z^2 = M_Z^2$, $\mu_b^2=m_b^2$, $\mu_\tau^2=m_\tau^2$ and $\mu_c^2=m_c^2$ up and down by
a factor of 2. 
The $\mu_b^2$ and $\mu_c^2$ variations are numerically small.
This perturbative uncertainty is added in quadrature to the
propagated error of SM inputs.
The total error is dominated by weak scale matching as estimated by $\mu_Z$ variation. 

\section{Low-energy leptonic theory}
  
\begin{table*}[h]
\centering
\caption{Effective couplings (in units $10^{-5}~\mathrm{GeV^{-2}}$) in
  the Fermi theory of neutrino-lepton scattering, 
  at renormalization scale $\mu=m_\mu$ (in the theory with neutrinos, $e$ and $\mu$)
  and in the low-energy limit at $\mu=m_e$ (in the theory with neutrinos and $e$).
  The error is dominated by the light-quark contribution.
}
\label{results_couplings_Running2}
\begin{minipage}{\linewidth}  
\footnotesize
\centering
\begin{tabular}{|l|c|c|c|c|c|c|c|}   
\hline          
&$ c^{\nu_\ell \ell'}_\mathrm{L},~\ell = \ell'$ & $ c^{\nu_\ell \ell'}_\mathrm{L},~\ell \neq \ell'$ & $ c^{\nu_\tau}_\mathrm{L}$ & $ c^{\nu_e}_\mathrm{R}$ & $c^{\nu_\mu}_\mathrm{R}$ & $c^{\nu_\tau}_\mathrm{R}$ \\
\hline
$ \mu = m_\mu$ &  $2.3997(29)$ &$-0.8994(29)$ & $-0.8921(29)$  & $0.7706(29)$  & $0.7706(29)$ & $0.7779(29)$ \\
\hline
$\mu = m_e $ & $2.3865(29) $ & $ -0.8988(29)$ & $-0.8916(29)$ & $0.7575(29) $ & $0.7711(29)$ & $0.7784(29)$\\
\hline
\end{tabular}
\end{minipage}
\end{table*}

For certain processes, the momentum transfer entering the loop of
Fig.~\ref{one_loop_ET} is below the scale where perturbative QCD is
applicable. 
An important example is 
the neutrino-electron scattering process where momentum transfer
is bounded by $Q^2 \le 2 m_e E_\nu \lesssim 0.01~\mathrm{GeV}^2$
for incoming neutrino energy $E_\nu \lesssim 10~\mathrm{GeV}$.
This process is described by an effective theory in which
hadrons and heavy leptons are integrated out and only neutrinos and electrons
remain as dynamical fields.  Here we determine the four-fermion couplings
in this low-energy leptonic theory. 
We perform the matching in a series of steps, first integrating out hadrons,
then muons. For the hadronic contribution to effective couplings, we 
employ results from Ref.~\cite{Tomalak:2019ibg}.
In a final step, we also describe 
the extreme low-energy limit applicable to forward scattering on nonrelativistic electrons.
Note that integrating out a heavy charged lepton violates the
equality of couplings $c^{\nu_\ell \ell^\prime}_\mathrm{R}$ for any $\ell$ and $\ell^\prime$, 
and of $c^{\nu_\ell \ell^\prime}_\mathrm{L}$ for any $\ell \ne \ell^\prime$. Consequently, 
the conventional definition of Weinberg angle in terms of effective couplings is flavor-independent 
only above the $\tau$-lepton mass scale.
Here we describe details for decoupling the $\mu$ lepton; a similar
procedure involving $\tau$ resulted in the last rows of
Table~\ref{results_couplings_running}.

In the theory valid below the hadron mass scale, i.e., $\mu \lesssim m_\pi$,
only neutrino couplings to muons and electrons contribute to running.
The renormalization group equations in the theory above the muon mass scale, 
i.e., $m_\mu \leq \mu \leq m_\pi $, are
given by
\begin{align}
  \frac{\mathrm{d} }{\mathrm{d} \ln \mu^2} \left( c^{\nu_\mu e}_\mathrm{L}  +  c_\mathrm{R}  \right)
  = \frac{\mathrm{d} }{\mathrm{d} \ln \mu^2} \left( c^{\nu_e e}_\mathrm{L}  +  c_\mathrm{R}  \right)
  = \beta_\ell \left( c^{\nu_\mu e}_\mathrm{L} +  c^{\nu_e e}_\mathrm{L} +  2 c_\mathrm{R}  \right) \,,
\quad 
  \frac{\mathrm{d} \ln \alpha  }{\mathrm{d} \ln \mu^2} = 2 \beta_\ell  \,, \label{alpha_running2} 
\end{align}
where $c^{\nu_e e}_\mathrm{R} = c^{\nu_\mu e}_\mathrm{R} = c_\mathrm{R}$.  The solution is
\ber
c_\mathrm{R} \left( \mu \right) = \frac{\alpha \left( \mu  \right)}{\alpha \left( \mu_0  \right)} c_\mathrm{R} \left( \mu_0 \right) + \left( 1 - \frac{\alpha \left( \mu  \right)}{\alpha \left( \mu_0  \right)}  \right)  \frac{ \mathrm{\tilde{G}}_e - \mathrm{G}_\mathrm{F}  }{\sqrt{2}}.
\eer
In the theory below the muon mass scale and above the electron mass scale,
i.e., $m_e \leq \mu \leq m_\mu$, RGEs for different neutrino flavors
decouple:
\begin{align}
  \frac{\mathrm{d} }{\mathrm{d} \ln \mu^2} \left( c^{\nu_\mu e}_\mathrm{L}   +  c^{\nu_\mu e}_\mathrm{R}   \right)
  = \beta_\ell  \left( c^{\nu_\mu e}_\mathrm{L}  +  c^{\nu_\mu e}_\mathrm{R}  \right) \,, \quad
  \frac{\mathrm{d} }{\mathrm{d} \ln \mu^2} \left( c^{\nu_e e}_\mathrm{L}  +  c^{\nu_e e}_\mathrm{R} \right)
  = \beta_\ell  \left( c^{\nu_e e}_\mathrm{L}   +  c^{\nu_e e}_\mathrm{R}  \right), \quad
\frac{\mathrm{d} \ln \alpha }{\mathrm{d} \ln \mu^2} = \beta_\ell \,,\label{alpha_running3}
\end{align}
resulting in distinct right-handed couplings in the scattering of electron- and muon-type
neutrinos:%
\footnote{It is convenient to perform the matching between effective theories with and without
  the muon degree of freedom at exactly $\mu= m_\mu$.  In this case, threshold  
  matching corrections to effective couplings and running $\alpha$
  vanish up to neglected corrections of relative order $\alpha^2$.}
\begin{align}
  c^{\nu_\mu e}_\mathrm{R} \left( \mu \right) &=
  \frac{\alpha \left( \mu  \right)}{\alpha \left( m_\mu  \right)} c_\mathrm{R} \left( m_\mu \right) + \left( 1 - \frac{\alpha \left( \mu  \right)}{\alpha \left( m_\mu  \right)} \right)  \frac{ \mathrm{\tilde{G}}_e}{\sqrt{2}}, \\
c^{\nu_e e}_\mathrm{R} \left(  \mu  \right) &=  \frac{\alpha \left( \mu  \right)}{\alpha \left( m_\mu  \right)} c_\mathrm{R} \left( m_\mu \right)  + \left( 1 - \frac{\alpha \left( \mu  \right)}{\alpha \left( m_\mu  \right)} \right) \frac{ \mathrm{\tilde{G}}_e - 2  \mathrm{G}_\mathrm{F}}{\sqrt{2}}.
\end{align}

For momentum transfers below the electron mass scale,
the effective theory in the one-electron sector
describes the interaction of neutrinos with a static electron source.\footnote{Corrections to the static limit may be described by nonrelativistic EFT.}
In the theory valid for $\mu \le m_e$, the electron is no longer involved in dynamics 
and the four-fermion couplings are scale invariant:
\ber
c^{\nu_\ell e}_{\mathrm{L},\mathrm{R}} \left( \mu \right) = c^{\nu_\ell e}_{\mathrm{L},\mathrm{R}} \left( m_e \right),
\qquad \ell = e, \mu, \tau \,. 
\eer
These effective couplings may be interpreted as effective radii,
analogous to the description of low momentum neutron scattering on charged particles~\cite{Koester:1995nx,Kopecky:1997rw}.
In detail, we may define the effective radius of interaction as
\begin{align} \label{eq:redef}
  - Q_e\frac{e_0^2}{6} r^2_{\nu_\ell e} &=
  \frac12 \left[  c_\mathrm{L}^{\nu_\ell e} (m_e)+ c_\mathrm{R}^{\nu_\ell e}(m_e) \right] \,, 
\end{align}
where $e_0^2 = 4\pi\alpha_0$ denotes the QED coupling in the low-energy limit. 
The matching from the $n_f=3$ or $n_f=4$ quark level theory to
the leptonic theory involves nonperturbative QCD.  Taking 
the relevant photon vacuum polarization tensor $\hat{\Pi}_{\gamma \gamma}(0)$
from experimental data, the result for electron neutrino effective radius is 
\begin{align}\label{eq:r2}
  r_{\nu_e e}^2 &= \bigg[ 40.05 + 0.36 \left( {\hat{\Pi}_{3\gamma}^{(3)}(0)\over \hat{\Pi}_{\gamma\gamma}^{(3)}(0)} - 1 \right)
    \bigg] \times 10^{-6}\,{\rm fm}^2
= ( 40.05 \pm 0.08 ) \times 10^{-6}\,{\rm fm}^2.
\end{align}
The uncertainty $0.08\,{\rm fm}^2$ results from
$\hat{\Pi}_{3\gamma}^{(3)}(0) / \hat{\Pi}_{\gamma\gamma}^{(3)}(0) = 1 \pm 0.2$ as in Ref.~\cite{Tomalak:2019ibg}. 

The difference between radii for different neutrino flavors, $\nu_\ell$ and $\nu_{\ell^\prime}$, 
is independent of the target particle $\mathrm{T}$,
when $\mathrm{T}$ is distinct from the charged lepton partner of the incident neutrinos.
Examples include $\nu_\ell = \nu_\mu$ and $\nu_{\ell^\prime}= \nu_\tau$ scattering 
on electrons, $\mathrm{T} = e$;
or any flavors $\nu_\ell$ and $\nu_{\ell^\prime}$ scattering on protons or neutrons. 
In these cases, the difference
$r^2_{\nu_\ell \mathrm{T}}- r^2_{\nu_\ell^\prime \mathrm{T}} \equiv r^2_{\nu_\ell}- r^2_{\nu_\ell^\prime}$ represents
a radius that is intrinsic to the neutrino species.\footnote{Our definition of radii, in terms of four Fermi operators of left-handed neutrinos, is independent of the neutrino mass sector, in particular whether right-handed neutrinos are introduced to form a Dirac mass, or a Majorana mass term is included.  For a general discussion of electromagnetic interactions of Majorana neutrinos, see Ref.~\cite{Kayser:1982br,Shrock:1982sc}.}
Starting from the theory with three active charged leptons (e.g. the
$n_f=4$ flavor theory renormalized at $\mu=2\,{\rm GeV}$ in Table~\ref{results_couplings_running}),
the differences in radii are induced by loops of charged leptons.
Summing over leptons in the loop and using the relations (\ref{eq:cldef}), we obtain%
\footnote{
  Our normalization for the lepton vacuum polarization function is as in Ref.~\cite{Tomalak:2019ibg}:
  $\Pi(0,m,\mu) = \mathrm{N_c} \left[ \frac13 \ln\frac{\mu^2}{m^2} + \frac{\alpha_s}{4\pi} \mathrm{C_F} \left( \ln\frac{\mu^2}{m^2} + \frac{15}{4} \right)\right]$,
    in terms of pole mass $m$.  For QED, we take  $\mathrm{N_c}\to 1$, $\mathrm{C_F}\to 1$, $\alpha_s \to \alpha$. We also have replaced $\alpha(\mu)$ by $\alpha_0$ including vacuum polarization, electron vertex and field renormalization corrections.}
\begin{align}\label{eq:rexpression}
  {e_0^2\over 6} (r^2_{\nu_\ell} -r^2_{\nu_{\ell'}} )
  &=- {\alpha_0 \over 2\pi} \left[ \Pi(0,m_\ell, \mu) - \Pi(0,m_{\ell^\prime},\mu) \right] c(\mu)
  = {\sqrt{2} \alpha_0 \GF \over 3\pi} \ln{m_\ell^2\over m_{\ell^\prime}^2} \left( 1 + {3\alpha(\mu) \over 4\pi} + \dots \right) \,.
\end{align}
In particular, 
\begin{align}
  r^2_{\nu_\mu} - r^2_{\nu_e} &= 3.476 \times 10^{-7}\,{\rm fm}^2
  \,, 
  \nl \label{eq:rmutau}
    r^2_{\nu_\tau} - r^2_{\nu_\mu} &= 1.840 \times 10^{-7}\,{\rm fm}^2
    \,.
\end{align}
For the special case of $\ell =\tau$ and $\ell^\prime = \mu$,
the same result is obtained from Eq.~(\ref{eq:redef}) upon substitution of low-energy
coefficients from Table~\ref{results_couplings_Running2}:
$r^2_{\nu_\tau} - r^2_{\nu_{\mu}}  = (3/4\pi\alpha_0)[ c_\rL^{\nu_\tau e}(m_e) + c_\rR^{\nu_\tau e}(m_e) - c_\rL^{\nu_\mu e}(m_e) - c_\rR^{\nu_\mu e}(m_e)  ]$.
For the evaluation using Table~\ref{results_couplings_Running2}, hadronic corrections 
cancel in the difference.  The evaluations differ by 
matching corrections of $\order(\alpha^2)$, present in Eq.~(\ref{eq:rexpression})
but omitted in the RG analysis; this difference impacts digits not displayed in Table~\ref{results_couplings_Running2}. 

The effective potential in matter can be expressed in terms of the effective
radii as a sum over all target particles
 \ber
 V^{\nu_\ell} = -\frac{e_0^2}{6}  \sum \limits_\mathrm{T} n_\mathrm{T}\, Q_\mathrm{T}\, r^2_{\nu_\ell \mathrm{T}},
 \eer
 with particle charge $Q_\mathrm{T}$ and number density $n_\mathrm{T}$.
 In a charge-neutral medium consisting of protons, neutrons and electrons,
 differences $ r_{\nu_\tau \mathrm{T}}^2-  r_{\nu_\mu \mathrm{T}}^2$
 enter with an opposite sign for positively and negatively charged particles
 resulting in $V^{\nu_\tau} = V^{\nu_\mu}$ up to
 corrections suppressed by powers $m_\tau^2/M_W^2$~\cite{Botella:1986wy}.
 Such differences would appear as corrections to the weak scale matching
 coefficients in (\ref{effective_Lagrangian_all}), and as $\order(\GF^2)$ corrections to forward scattering computed using (\ref{effective_Lagrangian_all}). 

\section{Summary}

We have determined the low-energy neutrino-fermion
effective field theory in $\overline{\mathrm{MS}}$ renormalization
scheme as the basis for neutrino scattering on electrons~\cite{Tomalak:2019ibg}, nucleons
and nuclei at sub-percent level.  
Electroweak scale coupling constants were determined by
matching to the Standard Model including complete one-loop
electroweak corrections, two-loop mixed QCD-electroweak corrections
in the lepton sector, and the $\gamma_5$ scheme dependence of effective operators. 
Among the eleven independent effective couplings, only two
depend on the scale.  Solving the renormalization group equation for these couplings,
we have determined all parameters in the
quark level effective Lagrangian with $n_f=3$ or $n_f=4$ quark flavors. 
A complete error budget due to parametric inputs and higher-order perturbative corrections
is presented. 
Using experimental data and  SU(3) flavor symmetry constraints, we have evaluated
hadronic contributions to determine 
the matching onto the low-energy theory involving leptons and the extreme
low-energy theory describing neutrino interactions with static electric charge
distributions. The hadronic correction provides the dominant source of uncertainty in low-energy
neutral-current interactions.
As a byproduct of our analysis, we revisited 
the comparison of the Fermi coupling constant $\mathrm{G}_\mathrm{F}$ evaluated 
at the electroweak scale to extractions from muon lifetime measurements; the comparison
shows only $0.9\sigma$ tension with $6\times 10^{-4}$
relative difference.

\section*{Acknowledgments}
This work supported by the U.S. Department of Energy, Office of
Science, Office of High Energy Physics, under Award Number
DE-SC0019095. Fermilab is operated by Fermi Research Alliance, LLC
under Contract No. DE-AC02-07CH11359 with the United States Department
of Energy. The work of O. Tomalak was supported in part by the
Visiting Scholars Award Program of the Universities Research
Association. O. Tomalak would like to acknowledge the Fermilab theory
group and the theory group of Institute for Nuclear Physics at
Johannes Gutenberg-Universit\"at Mainz for warm hospitality and
support. O. Tomalak is thankful to Kaushik Borah for useful
discussions regarding kinematics of scattering experiments, William
Jay for advice regarding the literature on semileptonic
operators and Mao Zeng for an inspiring question at RADCOR2019 conference. 
FeynCalc~\cite{Mertig:1990an,Shtabovenko:2016sxi},
LoopTools~\cite{Hahn:1998yk}, JaxoDraw~\cite{Binosi:2003yf} and
Mathematica~\cite{Mathematica} were extremely useful in
this work.

\appendix

\section{$\overline{\mathrm{MS}}$ masses of vector bosons and $\order(\alpha \alpha_s \GF)$ corrections}
\label{app:MSbar_from_observables}

To relate pole and $\overline{\mathrm{MS}}$ masses of vector bosons at some scale $\mu$, we consider renormalization of the
vector boson propagator $\mathrm{D}_{\mu \nu}$, using the example of the $Z$ boson as illustration:
\ber
\mathrm{D}^{\mu \nu} = \frac{i  \left( - g^{\mu \nu} + \left( 1 - \xi_Z \right) \frac{ q^\mu q^\nu }{q^2_Z - \xi^2_Z M^2_Z} \right)}{q^2_Z - M^2_Z} + \frac{i  \left( - g^{\mu \rho} + \left( 1 - \xi_Z \right) \frac{ q^\mu q^\rho }{q^2_Z - \xi^2_Z M^2_Z} \right)}{q^2_Z - M^2_Z} \left( - i \Sigma_{\rho \sigma} \right) \mathrm{D}^{\sigma \nu} ,
\eer
with vector boson self-energy
\ber
 \Sigma^Z_{\rho \sigma} = \left( g_{\rho \sigma} - \frac{q_\rho q_\sigma}{q^2} \right)  \Sigma^Z_T +  \frac{q_\rho q_\sigma}{q^2}  \Sigma^Z_L .
\eer
The pole mass is determined by the inverse propagator:
\ber
\mathrm{D}_{\mu \nu}^{-1} =  i g_{\mu \nu}  \left( q^2 - M^2_Z \right) + i \frac{1 - \xi_Z}{\xi_Z} q_\mu q_\nu + i \left( g_{\mu \nu} - \frac{q_\mu q_\nu}{q^2} \right)  \Sigma^Z_T +  i \frac{q_\mu q_\nu}{q^2}  \Sigma^Z_L.
\eer
We have cross checked expressions in terms of scalar integrals from
Refs.~\cite{Denner:1991kt,Hill:2014yka} in Feynman-'t Hooft gauge and
have verified the gauge independence of mass renormalization at one
loop when including Higgs tadpoles. For numerical evaluations, we
exploit analytical expressions for scalar integrals obtained following
Refs.~\cite{tHooft:1978jhc,Ellis:2007qk}, work in Feynman-'t Hooft
gauge and do not include Higgs tadpoles. We account for the leading
QCD corrections of
Refs.~\cite{Kniehl:1989yc,Degrassi:1990tu,Fanchiotti:1992tu,Djouadi:1993ss,Avdeev:1994db,Chetyrkin:1995ix}
representing the fermionic contribution to the following $\overline{\mathrm{MS}}$
differences (cf. Ref.~\cite{Fanchiotti:1992tu}) at renormalization scale $\mu$,
in terms of $\overline{\mathrm{MS}}$ mass of top quark:
\begin{align}\label{eq:sigmas}
  \frac{\hat{\Sigma}^W_T \left( \mu,~q^2 = M^2_W \right)}{M^2_W} - \frac{\hat{\Sigma}^W_T\left( \mu,~q^2 = 0 \right)}{M^2_W}
  &= \frac{\alpha}{2 \pi s^2_W} \left( B_0^{(f)} + \frac{\alpha_s}{\pi} B_\mathrm{QCD}^{(f)}\right) \,,  \\
  \frac{\hat{\Sigma}^W_T \left( \mu,~q^2 = M^2_W \right)}{M^2_W} - \frac{\hat{\Sigma}^Z_T\left( \mu,~q^2 = M^2_Z \right)}{M^2_Z}
  &= \frac{\alpha}{2 \pi s^2_W} \left( C_0^{(f)} + \frac{\alpha_s}{\pi} C_\mathrm{QCD}^{(f)}\right) \,. 
\end{align}
The normalization at $q^2=0$ is given by~\cite{Fanchiotti:1992tu,Djouadi:1993ss,Avdeev:1994db,Chetyrkin:1995ix}
\ber
\hat{\Sigma}^W_T\left( \mu,~q^2 = 0 \right) = \frac{3}{4} \frac{\alpha}{\pi} \frac{m^2_t}{s^2_W} \left[ -\frac{1}{4} - \frac{1}{2} \ln \frac{\mu^2}{m_t^2} + \frac{\alpha_s}{3 \pi} \left( -\frac{13}{8} + \zeta (2) -  \ln \frac{\mu^2}{m_t^2}  - \frac{3}{2}  \ln^2 \frac{\mu^2}{m_t^2} \right)\right].
\eer
The $\order(\alpha)$ correction is given by
\begin{align}\label{eq:leading}
  B_0^{(f)} &=2 \left( \ln \frac{\mu^2}{M_W^2} + \frac{5}{3}\right) - \frac{\ln \omega_t}{2} - \frac{\omega_t \left( 1 + 2\omega_t \right)}{8} - \frac{\left( \omega_t - 1 \right)^2}{2} \left( 1 + \frac{\omega_t}{2} \right) \ln \left( 1- \frac{1}{\omega_t}\right)\,,
  \nonumber \\
  C_0^{(f)} &= B_0^{(f)} - \frac{1}{2 c^2_W} \ln \frac{\mu^2}{M_Z^2} - \frac{5}{3 c^2_W} \left( \frac{7}{4} - \frac{10}{3} s^2_W + \frac{40}{9} s^4_W \right) + \frac{3 r_t}{2 c^2_W} \left(   \frac{3}{4} - \Lambda \left( r_t \right)\right)
  \nonumber \\
&- \frac{1}{8 c^2_W} \left[ 1 + \left( 1 - \frac{8}{3} s^2_W \right)^2 \right] \left[ 6 \ln \frac{\mu^2}{M_Z^2} - \ln r_t  - \frac{1}{3} + 2 \left( 1 + 2 r_t \right) \bigg( 1 - \Lambda \left( r_t \right) \bigg) \right] \,,
\end{align}
with $ \Lambda \left( r_t \right) = \sqrt{4 r_t - 1} \sin^{-1} \left(4 r_t \right)^{-1/2}$.
The $\order(\alpha \alpha_s)$ contribution is expressed in terms of $B_{\mathrm{QCD}}^{(f)}$ and $C_{\mathrm{QCD}}^{(f)}$:
\begin{align}\label{eq:next_to_leading}
  B_{\mathrm{QCD}}^{(f)} &= \ln \frac{\mu^2}{M_W^2} - 4 \zeta(3) + \frac{55}{12} + \frac{1}{2} \left[ \ln \frac{\mu^2}{m_t^2} - 4 \zeta(3) + \frac{55}{12} + 4 \omega_t \left( \mathrm{F}_1 \left(\frac{1}{\omega_t}\right) - \mathrm{F}_1\left(0\right)\right) \right]
  \nonumber \\
  &- \omega_t \left[ 1 + 2 \omega_t - 2 \left(1 - \omega_t^2 \right) \ln \left( 1- \frac{1}{\omega_t}\right) \right] \left( \frac{3}{4} \ln \frac{\mu^2}{m_t^2} + 1\right) \,,
  \\
  C_{\mathrm{QCD}}^{(f)} &= \frac{3}{2} \left[\ln \frac{\mu^2}{M_W^2} + \left( - 4 \zeta(3) + \frac{55}{12}\right) \left( 1-\frac{20}{9} s^2_W \right) \frac{s^2_W}{c^2_W} \right] - \frac{\ln \omega_t}{2} + 2 \omega_t  \mathrm{F}_1\left(\frac{1}{\omega_t} \right) - \frac{\omega_t}{2} \mathrm{A}_1\left(\frac{1}{4 r_t} \right) - \frac{\ln \frac{\mu^2}{m_t^2}}{8 c^2_W}
  \nonumber \\
  &-  \frac{ \left( 1 - \frac{8}{3} s^2_W \right)^2}{2 c^2_W} \left[ \frac{1}{4} \ln \frac{\mu^2}{m_t^2} + r_t \mathrm{V}_1 \left(\frac{1}{4 r_t}\right) \right]
  + \frac{1}{c^2_W} \left(  \frac{7}{3} s^2_W - \frac{22}{9} s^4_W -\frac{5}{4} \right) \ln \frac{\mu^2}{M^2_Z} \nonumber \\
  &+ 2 \omega_t \left[ \frac{16}{9} s^2_W \left( 4 c_W^2 - 1 \right) - \omega_t + \left(1 - \omega_t^2 \right) \ln \left( 1- \frac{1}{\omega_t}\right) \right] \left( \frac{3}{4} \ln \frac{\mu^2}{m_t^2} + 1 \right)
  \nonumber \\
&+\frac{4 \omega_t \Lambda \left( r_t \right)}{1 - 4 r_t} \left[ 2 r_t - 1 - \frac{32}{9} r_t s^2_W \left( 1 - 4 c^2_W \right) \right]  \left( \frac{3}{4} \ln \frac{\mu^2}{m_t^2} + 1 \right) \,,
\end{align}
with functions $\mathrm{F}_1(x),~\mathrm{V}_1(x)$ and $\mathrm{A}_1(x)$ from Ref.~\cite{Kniehl:1989yc} for an argument $ 0 \le x < 1$.
QCD corrections to $\gamma Z$ mixing may be treated similarly; the result is equivalent to
replacing $\ln {\mu^2}/{m^2_t}$ in Eq.~(\ref{eq:photong}) as
\ber
\ln \frac{\mu^2}{m^2_t} \to \ln \frac{\mu^2}{m^2_t} + \frac{\alpha_s}{\pi} \left( \frac{13}{12} - \ln \frac{\mu^2}{m^2_t} \right).
\eer

In the $\overline{\mathrm{MS}}$ renormalization scheme,
masses of vector bosons are related by $M_W \left( \mu \right) = M_Z \left( \mu \right) c_W  \left( \mu \right) $ 
which we exploit for the evaluation of $s_W(\mu)$.
To determine $M_W \left( \mu \right)$ and $M_Z \left( \mu \right)$, we solve a system of
equations:\footnote{For the second row in Tables~\ref{results_couplings} and~\ref{results_couplings_quark},
  we exploit $\left(M_W^{\rm p} \right)^2 = M^2_W \left( \mu \right) -  \Sigma_\mathrm{T}^W \left( \mu \right)$ instead of Eq.~(\ref{first_equation_for_MSbar}).}
\begin{align}
\mathrm{G}_\mathrm{F} &= \frac{\pi \alpha \left( \mu \right)}{\sqrt{2} M^2_W(\mu) s_W^2(\mu) }  g(\mu) \label{first_equation_for_MSbar} \, , \\
\left(M_Z^{\rm p} \right)^2 &=  M^2_Z \left( \mu \right) -  \Sigma_\mathrm{T}^Z \left( \mu \right)\, ,\label{second_equation_for_MSbar}
\end{align}
with $\overline{\mathrm{MS}}$ masses entering $g(\mu)$ and $\Sigma_\mathrm{T}^Z(\mu)$.
The pole mass of $Z$ boson $M^\mathrm{p}_Z$ is given in terms of the
center of the $Z$ peak as reported in PDG, $M^\mathrm{PDG}_Z$, and the inclusive width $\Gamma_Z$ as
\ber
\left(M^\mathrm{p}_Z \right)^2 +  \Gamma_Z^2 = \left(M_Z^{\rm PDG} \right)^2 \,.
\eer
We use the top quark pole mass, $m_t^{\mathrm{p}} = m_t^{\rm PDG}$, as input to  
determine the $\overline{{\mathrm{MS}}}$ mass as a function of renormalization scale $\mu$,
$m_t \left( \mu \right)$, 
according to Refs.~\cite{Bernreuther:1981sg,Larin:1994va,Chetyrkin:1997un,Fusaoka:1998vc,Marquard:2015qpa,Kataev:2015gvt}.
We do not consider renormalon effects which can cause an ambiguity in top quark $\overline{{\mathrm{MS}}}$
mass around $110~\mathrm{MeV}$~\cite{Beneke:2016cbu}. After this determination, we
evaluate $M_W \left( \mu \right)$ and $M_Z \left( \mu \right)$ solving Eqs.~(\ref{first_equation_for_MSbar})~and~(\ref{second_equation_for_MSbar}).

For our numerical analysis, 
we consider the top quark mass, Higgs boson mass, electromagnetic coupling constant, strong coupling constant,
Fermi constant and $Z$-boson mass as independent inputs and propagate errors of these parameters.
From the inputs of Eqs.~(\ref{eq:alphainput}), (\ref{eq:GFinput}) and (\ref{eq:heavymassinput}),
we obtain the following $\overline{\mathrm{MS}}$ masses in our renormalization scheme:
\begin{align}
M_Z(M_Z) &= 92.3499(82)~\mathrm{GeV} \,, \nl
M_W(M_Z) &= 80.961(8)~\mathrm{GeV} \,, \nl
m_t(M_Z) &= 170.9(4)~\mathrm{GeV} \,. \label{MSmass}
\end{align}
The Weinberg angle is determined from the ratio of $\overline{\mathrm{MS}}$ masses, $c_W(M_Z) = M_W(M_Z) / M_Z(M_Z)$,  as
\ber
s_W^2(M_Z) = 1 - \frac{M^2_W(M_Z)}{M^2_Z(M_Z)} = 0.23144\pm0.00003. \label{sW2}
\eer
In expressions (\ref{MSmass},~\ref{sW2}) above, we present only the error propagated from the uncertainty of the Standard Model parameters. 
The error of unaccounted high order perturbation theory is discussed in the main text.

\section{QED and QCD couplings}
\label{app:alpha}

In order to evaluate the expressions (\ref{eq:cldef}) and (\ref{eq:cdef}) at $\mu=M_Z$,
we require values for $M_W(\mu)$, $M_Z(\mu)$, $s_W(\mu)$ and $\alpha(\mu)$,
evaluated in the $\overline{\rm MS}$ scheme without tadpoles, with the full SM particle content.
We begin by deriving the necessary QED and QCD couplings from Ref.~\cite{Tanabashi:2018oca},
which were defined in the theory without top quark (cf. Refs.~\cite{Erler:1998sy,Erler:2004in}).
We use the same precision of running and matching as described in
Section~\ref{sec:inputs}.\footnoteref{note2}
The translation from the inputs (\ref{eq:alphainput}) results in 
\begin{align}
\alpha^{\rm SM}(M_Z)^{-1} = 128.120(10), \qquad \qquad  ~
\alpha_s^{\rm SM}(M_Z)  = 0.1176(16) .
\end{align}
For comparison, our inputs together with solution of RGEs and threshold matching conditions
yield 
$\left[ \alpha^{(4)}  \left( \mu = m_\tau \right) \right]^{-1} = 133.476(7)$, and 
$\alpha_s^{(4)}  \left( \mu = m_\tau \right) = 0.325(15)$.


\begin{thebibliography}{99}
  
\bibitem{Tanabashi:2018oca} 
  M.~Tanabashi {\it et al.} [Particle Data Group],
  Phys.\ Rev.\ D {\bf 98}, no. 3, 030001 (2018).


\bibitem{Maki:1962mu} 
  Z.~Maki, M.~Nakagawa and S.~Sakata,
  Prog.\ Theor.\ Phys.\  {\bf 28}, 870 (1962).


\bibitem{Gribov:1968kq} 
  V.~N.~Gribov and B.~Pontecorvo,
  Phys.\ Lett.\  {\bf 28B}, 493 (1969).


\bibitem{Bilenky:1978nj} 
  S.~M.~Bilenky and B.~Pontecorvo,
  Phys.\ Rept.\  {\bf 41}, 225 (1978).


\bibitem{Bilenky:1987ty} 
  S.~M.~Bilenky and S.~T.~Petcov,
  Rev.\ Mod.\ Phys.\  {\bf 59}, 671 (1987)
  Erratum: [Rev.\ Mod.\ Phys.\  {\bf 61}, 169 (1989)]
  Erratum: [Rev.\ Mod.\ Phys.\  {\bf 60}, 575 (1988)].

\bibitem{Meregaglia:2016vxf} 
  A.~Meregaglia,
  JINST {\bf 11}, no. 12, C12040 (2016).
  
\bibitem{Alvarez-Ruso:2017oui} 
  L.~Alvarez-Ruso {\it et al.},
  Prog.\ Part.\ Nucl.\ Phys.\  {\bf 100}, 1 (2018).
  
\bibitem{Duyang:2019prb} 
  H.~Duyang, B.~Guo, S.~R.~Mishra and R.~Petti,
  Phys.\ Lett.\ B {\bf 795}, 424 (2019).


\bibitem{Acciarri:2016crz} 
  R.~Acciarri {\it et al.} [DUNE Collaboration],
  arXiv:1601.05471 [physics.ins-det].
  
\bibitem{Abi:2020evt}
B.~Abi \textit{et al.} [DUNE],
[arXiv:2002.03005 [hep-ex]].
    
\bibitem{Hyper-Kamiokande:2016dsw} 
  [Hyper-Kamiokande Collaboration],
  KEK-PREPRINT-2016-21, ICRR-REPORT-701-2016-1.
  
\bibitem{Marshall:2019vdy} 
  C.~M.~Marshall, K.~S.~McFarland and C.~Wilkinson,
  Phys.\ Rev.\ D {\bf 101}, no. 3, 032002 (2020).


\bibitem{Park:2015eqa} 
  J.~Park {\it et al.} [MINERvA Collaboration],
  Phys.\ Rev.\ D {\bf 93}, no. 11, 112007 (2016).
  
\bibitem{Valencia:2019mkf} 
  E.~Valencia {\it et al.} [MINERvA Collaboration],
  Phys.\ Rev.\ D {\bf 100}, 092001 (2019).



\bibitem{Sirlin:1980nh} 
  A.~Sirlin,
  Phys.\ Rev.\ D {\bf 22}, 971 (1980).


\bibitem{Marciano:1980pb} 
  W.~J.~Marciano and A.~Sirlin,
  Phys.\ Rev.\ D {\bf 22}, 2695 (1980)
  Erratum: [Phys.\ Rev.\ D {\bf 31}, 213 (1985)].


\bibitem{Aoki:1980ix} 
  K.~i.~Aoki, Z.~Hioki, R.~Kawabe, M.~Konuma and T.~Muta,
  Prog.\ Theor.\ Phys.\  {\bf 65}, 1001 (1981).


\bibitem{Bohm:1986rj} 
  M.~Bohm, H.~Spiesberger and W.~Hollik,
  Fortsch.\ Phys.\  {\bf 34}, 687 (1986).


\bibitem{Hollik:1988ii} 
  W.~F.~L.~Hollik,
  Fortsch.\ Phys.\  {\bf 38}, 165 (1990).


\bibitem{Denner:1991kt} 
  A.~Denner,
  Fortsch.\ Phys.\  {\bf 41}, 307 (1993).


\bibitem{Sarantakos:1982bp} 
  S.~Sarantakos, A.~Sirlin and W.~J.~Marciano,
  Nucl.\ Phys.\ B {\bf 217}, 84 (1983).


\bibitem{Ram:1967zza} 
  M.~Ram,
  Phys.\ Rev.\  {\bf 155}, 1539 (1967).


\bibitem{Salomonson:1974ys} 
  P.~Salomonson and Y.~Ueda,
  Phys.\ Rev.\ D {\bf 11}, 2606 (1975).


\bibitem{Zhizhin:1975kv} 
  E.~D.~Zhizhin, R.~V.~Konoplich and Y.~P.~Nikitin,
  Izv.\ Vuz.\ Fiz.\  {\bf 1975}, no. 12, 82 (1975).


\bibitem{Byers:1979af} 
  N.~Byers, R.~Ruckl and A.~Yano,
  Physica A {\bf 96}, no. 1-2, 163 (1979).


\bibitem{Green:1980bd} 
  M.~Green and M.~J.~G.~Veltman,
  Nucl.\ Phys.\ B {\bf 169}, 137 (1980)
  Erratum: [Nucl.\ Phys.\ B {\bf 175}, 547 (1980)].


\bibitem{Green:1980uc} 
  M.~Green,
  J.\ Phys.\ G {\bf 7}, 1169 (1981).


\bibitem{Aoki:1981kq} 
  K.~i.~Aoki and Z.~Hioki,
  Prog.\ Theor.\ Phys.\  {\bf 66}, 2234 (1981).


\bibitem{Hioki:1981gi} 
  Z.~Hioki,
  Prog.\ Theor.\ Phys.\  {\bf 67}, 1165 (1982).


\bibitem{Bardin:1983yb} 
  D.~Y.~Bardin and V.~A.~Dokuchaeva,
  Nucl.\ Phys.\ B {\bf 246}, 221 (1984).


\bibitem{Bardin:1983zm} 
  D.~Y.~Bardin and V.~A.~Dokuchaeva,
  Sov.\ J.\ Nucl.\ Phys.\  {\bf 39}, 563 (1984)
  [Yad.\ Fiz.\  {\bf 39}, 888 (1984)].


\bibitem{Bardin:1985fg} 
  D.~Y.~Bardin and V.~A.~Dokuchaeva,
  Sov.\ J.\ Nucl.\ Phys.\  {\bf 43}, 975 (1986)
  [Yad.\ Fiz.\  {\bf 43}, 1513 (1986)].


\bibitem{Mourao:1989vb} 
  A.~M.~Mourao, L.~Bento and B.~K.~Kerimov,
  Phys.\ Lett.\ B {\bf 237}, 469 (1990).


\bibitem{Weber:1991kf} 
  A.~Weber and L.~M.~Sehgal,
  Nucl.\ Phys.\ B {\bf 359}, 262 (1991).


\bibitem{Buccella:1992xy} 
  F.~Buccella, C.~Gualdi, G.~Miele and P.~Santorelli,
  Nuovo Cim.\ B {\bf 107}, 1343 (1992).


\bibitem{Bernabeu:1994kw} 
  J.~Bernabeu, S.~M.~Bilenky, F.~J.~Botella and J.~Segura,
  Nucl.\ Phys.\ B {\bf 426}, 434 (1994).


\bibitem{Bahcall:1995mm} 
  J.~N.~Bahcall, M.~Kamionkowski and A.~Sirlin,
  Phys.\ Rev.\ D {\bf 51}, 6146 (1995).


\bibitem{Passera:2000ug} 
  M.~Passera,
  Phys.\ Rev.\ D {\bf 64}, 113002 (2001).


\bibitem{Akhmedov:2018wlf} 
  E.~Akhmedov, G.~Arcadi, M.~Lindner and S.~Vogl,
  JHEP {\bf 1810}, 045 (2018).


\bibitem{Weinberg:1967tq} 
  S.~Weinberg,
  Phys.\ Rev.\ Lett.\  {\bf 19}, 1264 (1967).


\bibitem{tHooft:1971ucy} 
  G.~'t Hooft,
  Phys.\ Lett.\  {\bf 37B}, 195 (1971).


\bibitem{Fermi:1934hr} 
  E.~Fermi,
  Z.\ Phys.\  {\bf 88}, 161 (1934).


\bibitem{Feynman:1958ty} 
  R.~P.~Feynman and M.~Gell-Mann,
  Phys.\ Rev.\  {\bf 109}, 193 (1958).


\bibitem{Sudarshan:1958vf} 
  E.~C.~G.~Sudarshan and R.~e.~Marshak,
  Phys.\ Rev.\  {\bf 109}, 1860 (1958).

\bibitem{Tomalak:2019ibg} 
  O.~Tomalak and R.~J.~Hill,
  Phys.\ Rev.\ D {\bf 101}, no. 3, 033006 (2020).
  
\bibitem{tHooft:1973mfk} 
  G.~'t Hooft,
  Nucl.\ Phys.\ B {\bf 61}, 455 (1973).
  
\bibitem{Cabibbo:1963yz}
  N.~Cabibbo,
  Phys.\ Rev.\ Lett.\  {\bf 10} (1963) 531.
  
\bibitem{Kobayashi:1973fv} 
  M.~Kobayashi and T.~Maskawa,
  Prog.\ Theor.\ Phys.\  {\bf 49}, 652 (1973).

\bibitem{Degrassi:1989ip} 
  G.~Degrassi, A.~Sirlin and W.~J.~Marciano,
  Phys.\ Rev.\ D {\bf 39}, 287 (1989).


\bibitem{Degrassi:1992ff} 
  G.~Degrassi and A.~Sirlin,
  Nucl.\ Phys.\ B {\bf 383}, 73 (1992).


\bibitem{Degrassi:1992ue} 
  G.~Degrassi and A.~Sirlin,
  Phys.\ Rev.\ D {\bf 46}, 3104 (1992).
  
\bibitem{Buras:1989xd} 
  A.~J.~Buras and P.~H.~Weisz,
  Nucl.\ Phys.\ B {\bf 333}, 66 (1990).
  
\bibitem{Dugan:1990df} 
  M.~J.~Dugan and B.~Grinstein,
  Phys.\ Lett.\ B {\bf 256}, 239 (1991).
  
\bibitem{Herrlich:1994kh} 
  S.~Herrlich and U.~Nierste,
  Nucl.\ Phys.\ B {\bf 455}, 39 (1995).

\bibitem{Erler:1998sy} 
  J.~Erler,
  Phys.\ Rev.\ D {\bf 59}, 054008 (1999).

\bibitem{Fanchiotti:1992tu} 
  S.~Fanchiotti, B.~A.~Kniehl and A.~Sirlin,
  Phys.\ Rev.\ D {\bf 48}, 307 (1993).
  
\bibitem{Kniehl:1989yc} 
  B.~A.~Kniehl,
  Nucl.\ Phys.\ B {\bf 347}, 86 (1990).

\bibitem{Chetyrkin:1995ix} 
  K.~G.~Chetyrkin, J.~H.~Kuhn and M.~Steinhauser,
  Phys.\ Lett.\ B {\bf 351}, 331 (1995).
  

\bibitem{Degrassi:1990tu} 
  G.~Degrassi, S.~Fanchiotti and A.~Sirlin,
  Nucl.\ Phys.\ B {\bf 351}, 49 (1991).

\bibitem{Avdeev:1994db} 
  L.~Avdeev, J.~Fleischer, S.~Mikhailov and O.~Tarasov,
  Phys.\ Lett.\ B {\bf 336}, 560 (1994)
  Erratum: [Phys.\ Lett.\ B {\bf 349}, 597 (1995)].
  
\bibitem{Djouadi:1993ss} 
  A.~Djouadi and P.~Gambino,
  Phys.\ Rev.\ D {\bf 49}, 3499 (1994).


\bibitem{Ward:1950xp} 
  J.~C.~Ward,
  Phys.\ Rev.\  {\bf 78}, 182 (1950).


\bibitem{Takahashi:1957xn} 
  Y.~Takahashi,
  Nuovo Cim.\  {\bf 6}, 371 (1957).


\bibitem{Aldins:1970id} 
  J.~Aldins, T.~Kinoshita, S.~J.~Brodsky and A.~J.~Dufner,
  Phys.\ Rev.\ D {\bf 1}, 2378 (1970).


\bibitem{Lautrup:1977tc} 
  B.~E.~Lautrup and M.~A.~Samuel,
  Phys.\ Lett.\  {\bf 72B}, 114 (1977).


\bibitem{Antonelli:1980zt} 
  F.~Antonelli and L.~Maiani,
  Nucl.\ Phys.\ B {\bf 186}, 269 (1981).


\bibitem{Arason:1991ic} 
  H.~Arason, D.~J.~Castano, B.~Keszthelyi, S.~Mikaelian, E.~J.~Piard, P.~Ramond and B.~D.~Wright,
  Phys.\ Rev.\ D {\bf 46}, 3945 (1992).


\bibitem{Erler:2004in} 
  J.~Erler and M.~J.~Ramsey-Musolf,
  Phys.\ Rev.\ D {\bf 72}, 073003 (2005).
  
\bibitem{Erler:2013xha} 
  J.~Erler and S.~Su,
  Prog.\ Part.\ Nucl.\ Phys.\  {\bf 71}, 119 (2013).


\bibitem{Erler:2017knj} 
  J.~Erler and R.~Ferro-Hern\'{a}ndez,
  JHEP {\bf 1803}, 196 (2018).
  
\bibitem{Kataev:1992dg} 
  A.~L.~Kataev,
  Phys.\ Lett.\ B {\bf 287}, 209 (1992).
  
\bibitem{Barbieri:1992dq} 
  R.~Barbieri, M.~Beccaria, P.~Ciafaloni, G.~Curci and A.~Vicere,
  Nucl.\ Phys.\ B {\bf 409}, 105 (1993).


\bibitem{Jegerlehner:2001fb} 
  F.~Jegerlehner, M.~Y.~Kalmykov and O.~Veretin,
  Nucl.\ Phys.\ B {\bf 641}, 285 (2002).


\bibitem{Jegerlehner:2002em} 
  F.~Jegerlehner, M.~Y.~Kalmykov and O.~Veretin,
  Nucl.\ Phys.\ B {\bf 658}, 49 (2003).


\bibitem{Degrassi:2014sxa} 
  G.~Degrassi, P.~Gambino and P.~P.~Giardino,
  JHEP {\bf 1505}, 154 (2015).


\bibitem{Martin:2015lxa} 
  S.~P.~Martin,
  Phys.\ Rev.\ D {\bf 91}, no. 11, 114003 (2015).


\bibitem{Martin:2015rea} 
  S.~P.~Martin,
  Phys.\ Rev.\ D {\bf 92}, no. 1, 014026 (2015).


\bibitem{vanRitbergen:1999fi} 
  T.~van Ritbergen and R.~G.~Stuart,
  Nucl.\ Phys.\ B {\bf 564}, 343 (2000).


\bibitem{Marciano:1999ih} 
  W.~J.~Marciano,
  Phys.\ Rev.\ D {\bf 60}, 093006 (1999).


\bibitem{Marciano:2011zz} 
  W.~J.~Marciano,
  J.\ Phys.\ Conf.\ Ser.\  {\bf 312}, 102002 (2011).


\bibitem{Barczyk:2007hp} 
  A.~Barczyk {\it et al.} [FAST Collaboration],
  Phys.\ Lett.\ B {\bf 663}, 172 (2008).


\bibitem{Casella:2013bla} 
  C.~Casella {\it et al.},
  Nucl.\ Instrum.\ Meth.\ A {\bf 700}, 1 (2013).


\bibitem{Webber:2010zf} 
  D.~M.~Webber {\it et al.} [MuLan Collaboration],
  Phys.\ Rev.\ Lett.\  {\bf 106}, 041803 (2011)
  [Phys.\ Rev.\ Lett.\  {\bf 106}, 079901 (2011)].


\bibitem{Tishchenko:2012ie} 
  V.~Tishchenko {\it et al.} [MuLan Collaboration],
  Phys.\ Rev.\ D {\bf 87}, no. 5, 052003 (2013).
  

\bibitem{Sirlin:1977sv} 
  A.~Sirlin,
  Rev.\ Mod.\ Phys.\  {\bf 50}, 573 (1978)
  Erratum: [Rev.\ Mod.\ Phys.\  {\bf 50}, 905 (1978)].
  
\bibitem{Altarelli:1980fi} 
  G.~Altarelli, G.~Curci, G.~Martinelli and S.~Petrarca,
  Nucl.\ Phys.\ B {\bf 187}, 461 (1981).

\bibitem{Sirlin:1981ie} 
  A.~Sirlin,
  Nucl.\ Phys.\ B {\bf 196}, 83 (1982).
  
\bibitem{Marciano:1985pd} 
  W.~J.~Marciano and A.~Sirlin,
  Phys.\ Rev.\ Lett.\  {\bf 56}, 22 (1986).
  
\bibitem{Marciano:1993sh} 
  W.~J.~Marciano and A.~Sirlin,
  Phys.\ Rev.\ Lett.\  {\bf 71}, 3629 (1993).
  
\bibitem{Erler:2002mv} 
  J.~Erler,
  Rev.\ Mex.\ Fis.\  {\bf 50}, 200 (2004).
  
  
\bibitem{Caswell:1974cj} 
  W.~E.~Caswell and F.~Wilczek,
  Phys.\ Lett.\  {\bf 49B}, 291 (1974).

\bibitem{Kopecky:1997rw} 
  S.~Kopecky, M.~Krenn, P.~Riehs, S.~Steiner, J.~A.~Harvey, N.~W.~Hill and M.~Pernicka,
  Phys.\ Rev.\ C {\bf 56}, 2229 (1997).

\bibitem{Koester:1995nx} 
  L.~Koester, W.~Waschkowski, L.~V.~Mitsyna, G.~S.~Samosvat, P.~Prokofevs and J.~Tambergs,
  Phys.\ Rev.\ C {\bf 51}, 3363 (1995).
  
\bibitem{Kayser:1982br}
B.~Kayser,
Phys. Rev. D \textbf{26}, 1662 (1982).

\bibitem{Shrock:1982sc}
R.~E.~Shrock,
Nucl. Phys. B \textbf{206}, 359-379 (1982).

  
\bibitem{Botella:1986wy} 
  F.~J.~Botella, C.~S.~Lim and W.~J.~Marciano,
  Phys.\ Rev.\ D {\bf 35}, 896 (1987).
  
\bibitem{Mertig:1990an} 
  R.~Mertig, M.~Bohm and A.~Denner,
  Comput.\ Phys.\ Commun.\  {\bf 64}, 345 (1991).
  
\bibitem{Shtabovenko:2016sxi} 
  V.~Shtabovenko, R.~Mertig and F.~Orellana,
  Comput.\ Phys.\ Commun.\  {\bf 207}, 432 (2016).
  
\bibitem{Hahn:1998yk} 
  T.~Hahn and M.~Perez-Victoria,
  Comput.\ Phys.\ Commun.\  {\bf 118}, 153 (1999).
  
\bibitem{Binosi:2003yf} 
  D.~Binosi and L.~Theussl,
  Comput.\ Phys.\ Commun.\  {\bf 161}, 76 (2004).
  
\bibitem{Mathematica} 
  W.~R.~Inc.,
  ``Mathematica, Version 11.0.1.0,''
   Champaign, IL, 2016.
  
\bibitem{Hill:2014yka} 
  R.~J.~Hill and M.~P.~Solon,
  Phys.\ Rev.\ D {\bf 91}, 043504 (2015).
  
\bibitem{tHooft:1978jhc} 
  G.~'t Hooft and M.~J.~G.~Veltman,
  Nucl.\ Phys.\ B {\bf 153}, 365 (1979).

\bibitem{Ellis:2007qk} 
  R.~K.~Ellis and G.~Zanderighi,
  JHEP {\bf 0802}, 002 (2008).
 
\bibitem{Bernreuther:1981sg} 
  W.~Bernreuther and W.~Wetzel,
  Nucl.\ Phys.\ B {\bf 197}, 228 (1982)
  Erratum: [Nucl.\ Phys.\ B {\bf 513}, 758 (1998)].
  
\bibitem{Larin:1994va} 
  S.~A.~Larin, T.~van Ritbergen and J.~A.~M.~Vermaseren,
  Nucl.\ Phys.\ B {\bf 438}, 278 (1995).

\bibitem{Chetyrkin:1997un} 
  K.~G.~Chetyrkin, B.~A.~Kniehl and M.~Steinhauser,
  Nucl.\ Phys.\ B {\bf 510}, 61 (1998).

\bibitem{Fusaoka:1998vc} 
  H.~Fusaoka and Y.~Koide,
  Phys.\ Rev.\ D {\bf 57}, 3986 (1998).


\bibitem{Marquard:2015qpa} 
  P.~Marquard, A.~V.~Smirnov, V.~A.~Smirnov and M.~Steinhauser,
  Phys.\ Rev.\ Lett.\  {\bf 114}, no. 14, 142002 (2015).
  
\bibitem{Kataev:2015gvt} 
  A.~L.~Kataev and V.~S.~Molokoedov,
  Eur.\ Phys.\ J.\ Plus {\bf 131}, no. 8, 271 (2016).

\bibitem{Beneke:2016cbu} 
  M.~Beneke, P.~Marquard, P.~Nason and M.~Steinhauser,
  Phys.\ Lett.\ B {\bf 775}, 63 (2017).
  


\end{thebibliography}
\end{document}